\documentclass[5p,times]{elsarticle}
\usepackage{amssymb, amsmath, amsthm}
\usepackage{graphicx}
\usepackage{multirow}
\usepackage{algorithm}
\usepackage[noend]{algorithmic}
\usepackage{mathtools}
\usepackage{color,url} 
\usepackage[font=normalsize,labelfont={bf}, skip=3pt]{caption}

\usepackage{empheq}
\usepackage[most]{tcolorbox}
\newcommand{\et}{{\it et al.}}

\newcommand{\rev}[1]{{\color{black}{#1}}}
\newcommand{\duy}[1]{{\color{black}{#1}}}
\newcommand{\hg}[1]{{\color{black}{#1}}}

\theoremstyle{definition}
\newtheorem{definition}{Definition}

\journal{Future Generation Computer Systems}

\begin{document}

\begin{frontmatter}



\title{From Programming Bugs to Multimillion-Dollar Scams:\\An Analysis of Trapdoor Tokens on Uniswap} 

%

\author[rmit]{Phuong Duy Huynh\corref{cor1}}
\ead{duyhuynhdev@gmail.com}
\cortext[cor1]{Corresponding author at: School of Computing Technologies, RMIT University, 124 La Trobe St, Melbourne VIC 3000, Australia}
\author[rmit]{Son Hoang Dau}
\author[rmit]{Xiaodong Li}
\author[rmit]{Iqbal Gondal}
\author[rmit]{Thisal De Silva}
\author[monash]{Emanuele Viterbo}
\affiliation[rmit]{organization={School of Computing Technologies, RMIT University},
            city={Melbourne},
            postcode={3100}, 
            state={VIC},
            country={Australia}}
\affiliation[monash]{organization={Electrical and Computer Systems Eng. Dept., Monash University},
            city={Clayton},
            postcode={3800}, 
            state={VIC},
            country={Australia}}
\begin{abstract}
We investigate in this work a recently emerged type of scam ERC-20 token called Trapdoor, which has cost investors billions of US dollars on Uniswap, the largest decentralised exchange on Ethereum, from 2020 to 2023. In essence, Trapdoor tokens allow users to buy but preventing them from selling by embedding logical bugs and/or owner-only features in their smart contracts. 
By manually inspecting a number of Trapdoor samples,
we established the first systematic classification of Trapdoor tokens and a comprehensive list of techniques that scammers used to embed and conceal malicious codes, accompanied by a detailed analysis of representative scam contracts. 
\hg{In particular, we developed TrapdoorAnalyser, a fine-grained detection tool that generates and crosschecks the error-log of a buy-and-sell test and the list of embedded Trapdoor indicators from a contract-semantic check to reliably identify a Trapdoor token. 
TrapdoorAnalyser not only outperforms the state-of-the-art commercial tool GoPlus in accuracy, but also 
provides traces of malicious code with a full explanation, which most of the existing tools lack. 
Using TrapdoorAnalyser, we constructed the very first dataset of about \rev{30,000 Trapdoor and non-Trapdoor} tokens on UniswapV2, which allows us to train several machine learning algorithms that can detect with very high accuracy even Trapdoor tokens with no available Solidity source codes.} 
\end{abstract}

\begin{keyword}
Smart contracts \sep Ethereum \sep Trapdoor \sep Decentralised Exchanges \sep Cryptocurrencies \sep Crypto-Scams
\end{keyword}

\end{frontmatter}


\section{Introduction}
The widespread adoption of the blockchain technology together with the ever-growing demand for trading digital assets has led to the emergence of hundreds of cryptocurrency \textit{centralised exchanges} (CEXs) such as Binance~\cite{Binance2017}, Coinbase~\cite{CoinBase2012}, and KuCoin~\cite{KuCoin2017}, all of which employ the traditional trading mechanism with the vital role of a central authority. By contrast, \textit{decentralised exchanges} (DEXs) such as Uniswap~\cite{Uniswap}, Sushiswap~\cite{Sushiswap}, and Pancakeswap~\cite{Pancakeswap}, 
have been developed to facilitate decentralisation and enhance user privacy.
Governed solely by a set of smart contracts, DEXs allow users to trade their digital assets directly to each other \textit{without} any intermediary, hence providing them with full control of their assets, better anonymity, as well as censorship resistance. On the other hand, the lack of central authority on DEXs also means little quality control, regulations, and customer support, making their users susceptible to a plethora of issues, most notably price slippage, smart contracts bugs and vulnerabilities, and low-quality or downright malicious scam tokens~\cite{Blocktelegraph_CEX_DEX,Coindesk_CEX_DEX,mazorra2022not,Xia:2021}.

Founded in 2018, Uniswap has become one of the most popular DEXs operating on the Ethereum blockchain, with the daily trading volume exceeding US\$1.2 billion~\cite{Coingecko2023DEX} at the time of writing. However, Uniswap has also been reported to be littered with scam tokens~\cite{Xia:2021,mazorra2022not,cernera2023token}.
In a recent significant cryptocurrency-related court case~\cite{YahooNews_CourtCaseUniswap_2023,Memorendom_CourtCaseUniswap_2023}, six investors from North Carolina, Idaho, New York, and Australia, who lost money after investing in various scam tokens on Uniswap, decided to sue Universal Navigation Inc., the company behind the exchange, and its founder/CEO Hayden Z. Adams to the US District Court of New York. Uniswap's counsel argued that making them liable for scam tokens on their DEX is like holding ``a developer of self-driving cars liable for
a third party’s use of the car to commit a traffic violation or to rob a bank'', to which the judge agreed. The case was dismissed by the judge in August 2023 but has undoubtedly become a legal landmark on cryptocurrency investment scams.

The emergence of scams on Uniswap has drawn 
attention from the research community. 
Xia~\emph{et al.}~\cite{Xia:2021} were the first to characterise and measure scam tokens on this DEX, identifying over 10,000 scam tokens and related exchange pools as part of their work. They then analysed these scam tokens' behaviours and pointed out that  a \textit{Rug-pull} scam, in which the project developers first lured the investors into buying a new and seemingly profitable token and then disappeared with all the funds, is the most popular scam 
on Uniswap. 
Rug pull was also reported in the Crypto Crime Report from Chainalysis~\cite{Chainanalysis:2022} as the most common cryptocurrency financial scam and responsible
for 37\% of all cryptocurrency scam revenue in 2021. Although 
the work of Xia~\et~\cite{Xia:2021} 
provided a great exposition and overview of the Rug-pull scam, their detection approach is simple: they can only detect Rug-pull tokens after they have occurred and fail to 
detect more advanced types of Rug-pull scam like Trapdoor. Mazorra~\emph{et al.}~\cite{mazorra2022not} designed an automated Rug pull detection tool to predict future Rug-pull scams based on the Herfindahl–Hirschman Index (HHI) and clustering transaction coefficient heuristics. 
Furthermore, they expanded the Rug-pull dataset to 27,588 tokens and 
classified Rug pulls into three types:
simple, sell, and Trapdoor. However, their detection tool cannot distinguish each type, especially the Trapdoor.  In fact, Rug-pull is an umbrella term that refers to scams in which the project developers disappear with all of the funds, leaving the victims with worthless assets. Trapdoor, on the other hand, refers to a particular set of techniques that ensure that investors cannot sell back the tokens they have bought, which is a specific way to allow the pool to be rug-pulled later.

A Trapdoor token employs programming logical bugs such as an ``if'' condition that is never satisfied, a fee-manipulation mechanism that can only be called by the contract owner, or numerical exceptions such as division by zero, in order to allow the investors to buy newly created tokens but prevent them from selling the tokens back to the pool to earn a profit. Despite the fact that Trapdoor tokens have been noticed by both research community and 
industry~\cite{Xia:2021,mazorra2022not,coinbrain_post,cryptobriefing_post,quillaudits_post,rampiro_post}, 
a systematic and comprehensive study of Trapdoor tokens on Uniswap is still missing. We seek to address this research gap. Our main contributions are discussed below.
\textbf{A systematic and comprehensive analysis of Trapdoor tokens.} As far as we know, this study is the first work that conducts a systematic and comprehensive analysis of the sophisticated Trapdoor tokens on Uniswap, 
including demonstrations of how they work with real-world examples (Section~\ref{subsec:Trapdoor_classification}). Especially, we categorise five different types of Trapdoor techniques according to their intended use in trapping investors' fund. Moreover, we also 
classify different maneuvers that scammers use to 
hide their ``trap'' in token contracts. 
Such Trapdoor techniques and maneuvers can be used individually or in combination to create more sophisticated scam tokens (Section~\ref{subsec:Trapdoor_technique}).


\textbf{A new ground-truth Trapdoor dataset and a reliable  detection tool.} \hg{We propose TrapdoorAnalyser, a novel detection tool that can reliably identify a Trapdoor token by performing a simultaneous analysis (with cross-validation) of the error-log produced by a buy-and-sell test and the Trapdoor-indicator list generated by a contract-semantic check (Section~\ref{sec:Trapdoor_dataset}). 
TrapdoorAnalyser achieves better detection performance and is capable of identifying more advanced 
Trapdoor techniques compared to GoPlus~\cite{goplus}, a state-of-the-art detection tool from the industry.
Applying TrapdoorAnalyser on all UniswapV2 tokens from its launch to January 2023, we obtain a ground-truth dataset (available at \cite{trapdoor_data}) of 11,943 Trapdoor tokens and 18,548 non-Trapdoor tokens, together with their pools (Section~\ref{sec:labelling}).}


\textbf{Accurate machine-learning Trapdoor detection models.} 
Our aforementioned approach only works on tokens that have smart contract source codes available. According to our data analysis (Fig.~\ref{fig:token_and_exchange_pool}, Section~\ref{subsec:trapdoor_samples}), about 23\% of ERC-20 tokens on UniswapV2 do \textit{not} have source codes available. To accommodate such tokens, we build machine-learning based detection models using features extracted from always-available data including transactions and operation codes (opcode) of token contracts. Experimental results indicate that this approach can help detect Trapdoor tokens and malicious techniques embedded in a token contract with very high F1 scores (Section~\ref{sec:ML_based_detection}). The Python code of our machine-learning based tool is available at~\cite{machine_learning_tool}.

\textbf{Scam impact analysis.} 
We systematically analyse Trapdoor tokens' characteristics, scammer tactics, and their financial impacts 
in  Section~\ref{sec:trapdoor_analysis}. According to our analysis, the Trapdoor scams on Uniswap has cost investors a billion US dollars between 2020 and 2023
. We also found in our analysis that around 267,000 investors (unique addresses) had bought such tokens. 
Furthermore, our analysis also reports tactics that scammers use to mislead investors, 
including generating fake tokens \hg{having similar or the same names as well-known tokens or companies} to create a false sense of trustworthiness or employing multiple clones of the same Trapdoor tokens to \hg{quickly and cheaply scale up their scam activities}. 

\section{Background}
\label{sec:background}
\subsection{Ethereum Smart Contract and ERC-20 Token}
\textit{Smart contract}. The concept of a smart contract, an automated executable program, was introduced in 1997 by Nick Szabo~\cite{NSzaboSmartContract1996}. However, this concept only got its first practical implementation in the release of Ethereum in 2015. Smart contracts on Ethereum can be implemented using a Turing-complete programming language called Solidity~\cite{Solidity}. A contract's source code is then compiled into \rev{a} low-level \textit{bytecode} and deployed onto the Ethereum blockchain through a transaction. Once the contract's \textit{bytecode} is stored successfully on the chain, it becomes immutable and publicly available on the chain. 
A unique address 
is provided to identify a newly deployed smart contract, which users can use later to interact with the contract. Any communication with the contract, e.g., function invoking, will be recorded in its transaction history with complete information about the function called, function input, execution time, etc.

\textit{ERC-20 Token}. Fungible (interchangeable) tokens are 
smart contracts that can represent virtual assets, such as company shares, lottery tickets, or even fiat currencies. 
\hg{ERC-20 tokens are fungible tokens that follow
the \texttt{ERC-20 standard}~\cite{ERC20Standard}, which defines a set of rules 
including standard methods and events. 
In this work, we focus on \texttt{transfer} and \texttt{transferFrom}, 
the two fundamental methods used to transfer ERC-20 tokens between two arbitrary accounts. Moreover, 
Uniswap 
also relies on these two transfer functions to swap tokens back and forth. Therefore, scammers often modify 
these two functions or relevant functions called from them to implement their scam logic.} 


\subsection{Decentralised Exchange}
\label{subsec:dex}
Decentralised exchanges (DEXs) allow users to exchange their digital assets without the involvement of central authorities~\cite{werner2022sok}. 
More specifically, DEXs execute a digital trade without storing user funds (noncustodial exchange) or personal data on a centralised server. 
DEXs \rev{primarily operate} based on one of the two 
trading price determination mechanisms: \textit{order book} and \textit{automated market maker} (AMM). 
\rev{Similar to} the traditional stock exchanges, an order-book-based DEX performs a trade by recording traders' orders into the order book and waiting until the DEX finds a suitable order that matches the preset price, allowing traders to buy or sell their digital asset at the expected price. Unlike the order-book model, the price of \rev{a} digital asset in the AMM model is calculated using a mathematical formula. This model is more \rev{widely} 
adopted in blockchain environments due to its computational performance~\cite{werner2022sok}. In general, AMM DEXs work on the concept of a \textit{liquidity pool},  
in which users exchange assets by transferring one asset into the pool and withdrawing another following a calculable exchange 
ratio. The ratio effectively determines the price of an asset in the pool. Different DEXs use different formulas for such exchange ratios. 

\subsection{Token Exchange on Uniswap}
\label{subsec:exchange_flow}
\textit{Uniswap}~\cite{Uniswap}. Uniswap is the largest decentralised exchange that adopts the AMM model successfully. \rev{Uniswap} 
was launched onto the Ethereum network in 2018, and currently, 
has three different versions that operate independently. Among these three versions, UniswapV2 
outperforms other versions in terms of the number of listed tokens and the number of exchange pools (liquidity pools)~\cite{Coingecko2023DEX}. Moreover, among all DEXs in the cryptocurrency market, UniswapV2 is the top protocol forked by other DEXs (632 forks)~\cite{UniswapV2Folks} across different blockchains due to its popularity and availability. 
\rev{As such, we chose to study UniswapV2, while noting that our approach is also applicable to other versions and forks.} 

\begin{figure}[ht]
\centering
\includegraphics[width=0.4\textwidth]{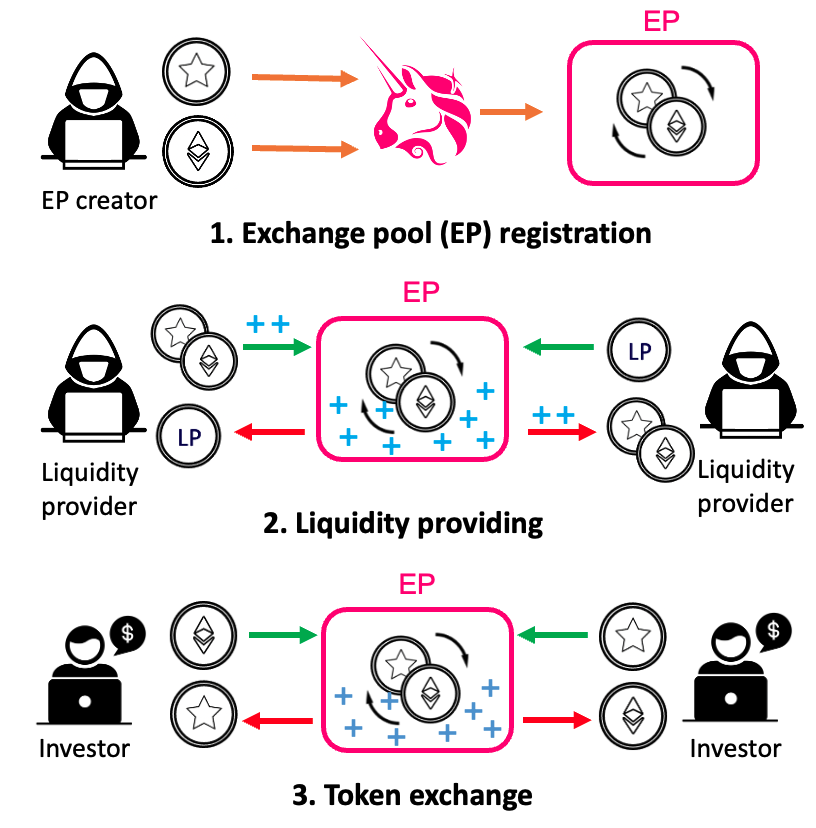}
\caption{Interaction flow in a UniswapV2 exchange pool.}\label{fig:uniswap_interaction_flow}
\end{figure}

\textit{Liquidity Pool.}
In UniswapV2, each liquidity pool resembles 
an automatic exchange counter for a pair of two different ERC-20 tokens. 
The interaction flow between users and a UniswapV2 pool is depicted in Fig.~\ref{fig:uniswap_interaction_flow} and further explained below.
Note that UniswapV2 operations are based on 
three main smart contracts: \texttt{Factory}, \texttt{Pair}, and \texttt{Router}. 

\textit{Exchange Pool (EP) Registration.} 
To create and launch a new exchange pool, a user must call the function \texttt{createPair} in the \texttt{Factory} contract with the two corresponding token addresses as input. The user who creates a pool is referred to as an exchange pool creator (EP creator).

\textit{Liquidity Providing.} After the liquidity pool is \rev{successfully launched}, any user (\rev{including} EP creator) will be able to deposit two paired tokens (liquidity) into the pool.
A user who added liquidity into the pool is called a \textit{liquidity provider} (LP).
Since a \texttt{Pair} itself is also an ERC-20 token, the pool \textit{mints} LP-tokens as the proof of liquidity contribution and 
sends back to the contributor \rev{whenever} it receives new liquidity. These LP tokens can later be \textit{burned} by the liquidity provider to withdraw the fund. In \rev{the} \texttt{Pair} contract, two corresponding functions \texttt{mint} and \texttt{burn} are provided to support the features mentioned above. 

\textit{Token Exchange.} When investors want to exchange one of the tokens in the pair for another, they must call the \texttt{swap} function of a liquidity pool. The exchange rate is calculated based on the ratio of tokens available in the pool, following the constant-product formula 
         $R_xR_y=\big(R_x+\Delta_x(1-\gamma)\big)(R_y-\Delta_y)$,
where $R_x$ and $R_y$ are the current amounts (reserves) of the two tokens in the pool, \rev{$\Delta_x$ and $\Delta_y$ are the amounts of tokens that the user deposits into and receives from the pool, respectively.} 
For each exchange, $\gamma = 0.003$ is the fee charged from the trader and proportionally distributed to all liquidity providers in the pool as rewards for their liquidity contribution. Hence, when users exchange one token for the other, the price of the latter will rise \duy{with updated reserves $(R_x \leftarrow R_x + \Delta_x, R_y\leftarrow R_y - \Delta_y)$}. In a swap, a user must first transfer his owned token to the pool. Then, the corresponding amount of the target token will be calculated based on the exchange rate and transferred to the buyer. 
 
 A newly created token is usually paired with a popular and high-value token to make the pool \rev{more attractive to} 
 traders. 
 \rev{Swapping a high-value token for a newly created token A is called \textit{buying A}, while the reverse is referred to as \textit{selling A}.}
 
\section{Trapdoor Tokens on UniswapV2}
\label{sec:Trapdoor_token}
\duy{In this section, we first provide a theoretical definition of Trapdoor and distinguish it 
from Honeypot. 
\hg{We then describe our collection process of all the ERC-20 tokens on UniswapV2 from May 2020 to January 2023 and report useful statistics. 
From this token dataset, we manually inspected 2,723 suspicious tokens with available source codes and obtained our first dataset of 1,859 Trapdoor tokens.
Note that to confirm if a token is a Trapdoor, we examined both its source code and the transaction history of the creator to see if the scam in fact went through. 
Based on a careful study of these tokens,} we identify and categorise malicious techniques that have been used to develop Trapdoor tokens. 
Finally, we provide a list of maneuvers that scammers use to \hg{embed} 
and conceal their malicious codes.}

\subsection{Definition of Trapdoor Tokens}
\label{subsec:Trapdoor_definition}

\begin{definition}[High-value Token]
\label{def:high_value}
A high-value token is a token that has a consistently high market cap and has been paired with many other tokens on exchanges. 
\hg{More precisely, we take a conservative approach and define a high-value token as a token with a market cap of at least a million dollars and that has been paired in at least 50 exchange pools.}
\end{definition}


\begin{definition}[Trapdoor Token]
\label{def:trapdoor}
\hg{A Trapdoor token is a token with a malicious contract code that allows investors to buy (by paying with a high-value token) but prevents them from gaining expected profit (in the form of a high-value token).}
\end{definition}

\begin{figure}[ht]
\centering
\includegraphics[width=0.4\textwidth]{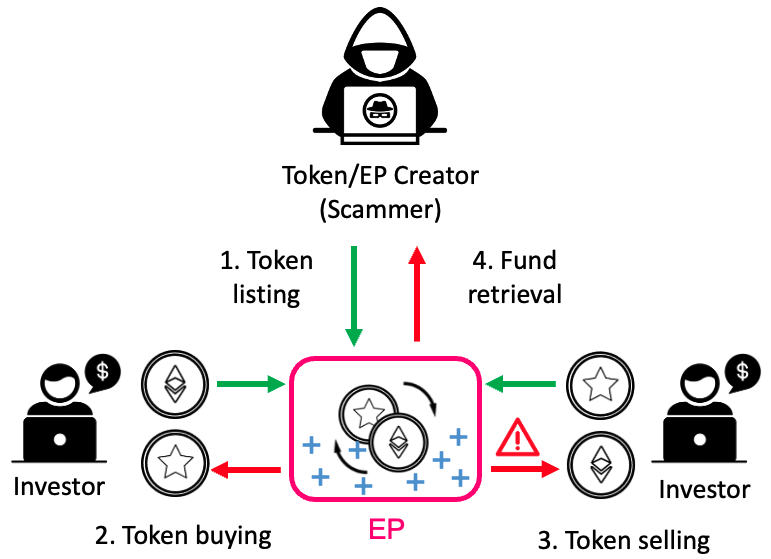}
\caption{The execution of a Trapdoor scam in four steps. An investor can buy the scam token from the exchange pool using a high-value token but cannot sell it back.} 
\label{fig:Trapdoor_process}
\vspace{-15pt}
\end{figure}

A Trapdoor scam on a DEX runs in four steps (see Fig.~\ref{fig:Trapdoor_process}).

\begin{enumerate}
    \item \textit{Token listing.} The scammer lists a Trapdoor token onto the chain and creates an exchange pool on DEX that \rev{pairs} 
    the Trapdoor token \rev{with} 
    a high-value token. In most cases, a Trapdoor token is paired with (wrapped) \texttt{ETH}.

    \item \textit{Token buying.} Once the exchange pool is available on DEX, the investors can buy the Trapdoor token by transferring the high-value token to the pool. 

    \item \textit{Token selling.} As more investors buy the Trapdoor token and add the high-value token to the pool, the value of the scam token rises with respect to the high-value token. However, the investors cannot sell to gain profit as they expected due to embedded \duy{malicious codes}. 
   
    \item \textit{Fund retrieval (Rug-pull).} The scammer/token creator withdraws all tokens from the exchange pool, including high-value tokens investors have invested and disappears.
\end{enumerate}

\vspace{-5pt}

\subsection{Trapdoor and Honeypot}
\label{subsec:Trapdoor_vs_honeypot}
Trapdoor tokens are sometimes referred to as Honeypot scams in several \rev{online detection tools}  
\cite{honeypotis,detecthoneypot,tokensniffer}. However, these two \rev{scams are fundamentally} 
different in \rev{many} 
aspects.

\textbf{Targeted victim.} Honeypot scams often target investigators with some level of experience who can read smart contracts, while Trapdoor scams target novice investors \hg{or trading bots} who cannot understand the contract very well.

\textbf{Scam technique.} Honeypot scams lure the victims by intentionally exposing an easy-to-spot loophole that seemingly could be exploited by investors to gain a big profit from the contract, which turns out to be a \textit{fake loophole}: the hopeful investor observes the (fake) loophole, invests into the contract, and ends up losing their investment. A Trapdoor token, on the contrary, aims to hide the malicious/buggy code, making it harder for users to detect. The techniques used in a Honeypot, as reported in~\cite{torres2019art}, are also very different from those in a Trapdoor.

\textbf{Ethical intention.} In Trapdoor scams, a token creator commits a severe ethical violation by intentionally stealing funds from investors who are actually the victims of this scam. However, in Honeypot scams, the ethical intentions of users are not explicit~\cite{torres2019art}. Although it is clear that a Honeypot creator intentionally deploys a contract to make a profit from users, users are only benign if they accidentally invest in a Honeypot contract. In most cases, users attempt to intentionally exploit vulnerabilities of a Honeypot contract and attack it to get profit from the \duy{creator}. 
\rev{Therefore, it is not clear whether Honeypot users are always benign.}

\subsection{\hg{Token and Pool Datasets on UniswapV2}}
\label{subsec:tokens_pools_datasets}

To collect samples of Trapdoor tokens \hg{(and also for building the full Trapdoor dataset in Section~\ref{sec:Trapdoor_dataset})}, we first gathered all the token addresses listed on UniswapV2 from May 2020, when this platform was launched, to January 2023, the time \rev{our study started.} 
We first queried from the Ethereum chain all the \texttt{PairCreated} events of \texttt{Factory} using web3 library~\cite{Web3Python}. These events store the creation information of generated liquidity pools on UniswapV2. As a result, we collected \hg{a pool dataset} of \textbf{137,454} liquidity pools along with \hg{a token dataset} of \textbf{131,172} unique token addresses listed on this platform. Besides, we also retrieved token information, including token names and symbols, creator addresses, and Solidity source codes via 
Etherscan APIs~\cite{EtherscanAPI} for further analyses.

\begin{figure}[ht]
\centering
\includegraphics[width=0.4\textwidth]{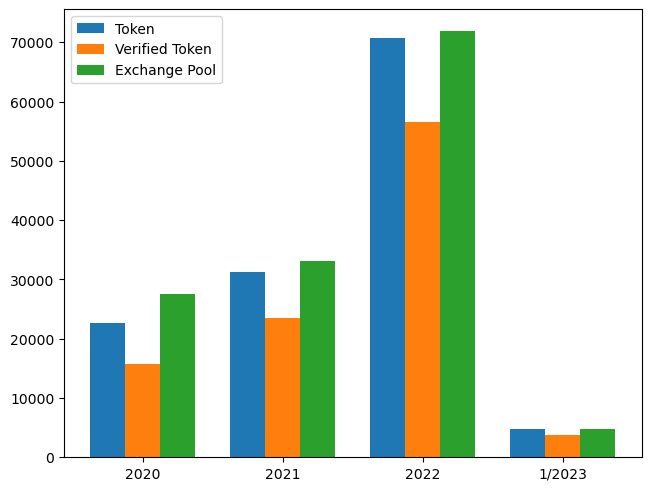}
\caption{The number of tokens and pools on UniswapV2.
}
\label{fig:token_and_exchange_pool}
\end{figure}

We demonstrate in Fig.~\ref{fig:token_and_exchange_pool} the yearly token listing and exchange pool creation events on UniswapV2 from our analysis. 
We observe that after the launch, UniswapV2 grew very quickly. 
In particular, 22,707 tokens were listed, and 27,543 corresponding exchange pools were created within seven months. The numbers of tokens and pools went up to 31,184 and 33,141 in 2021 and increased sharply 
to 70,693 and 71,961 in 2022, respectively. 
Among 131,172 listed tokens, about 77\% (101,050 tokens) tokens were verified with publically available contract source codes. The rest do not have source codes available. 

\subsection{Collecting Samples of Trapdoor Tokens}
\label{subsec:trapdoor_samples}
\hg{To study Trapdoor tokens and their characteristics, we need to find a sufficient number of Trapdoor samples and examine them carefully.
As it is impossible to have an accurate Trapdoor filter at this point, we look for \textit{suspicious} tokens that are likely to be Trapdoor, and then manually inspect them to find out the real Trapdoor tokens. 

\begin{tcolorbox}[colback=black!1,colframe=black]
 \textbf{R1} The token is paired with a high-value token.\\
 \textbf{R2} The token was sold by at most one holder.\\
 \textbf{R3} The token has been bought by at least ten investors.\\
 \textbf{R4} The token has been listed on DEX for at least one month at the data collection time.
\end{tcolorbox}

More specifically, a token is marked as \textit{suspicious} if it satisfies \textit{all} four requirements listed above. Note that these four conditions R1-R4 roughly capture our intuition of what a Trapdoor token should be like. There is no need to be highly accurate at this point (which is also an impossible task), as we will inspect them again for the actual Trapdoor tokens.} 
Indeed, 
\textbf{R1} dictates that scammers most likely will pair their tokens with a high-value token to attract investors, while 
\textbf{R2} requires that nobody can sell a Trapdoor token except for the token creator. 
Moreover, \textbf{R3} ensures that the token is successful in attracting investors and hence eliminating underperforming tokens. 
We use \textbf{R4} 
to filter out newly listed tokens that might behave like Trapdoor regarding the transaction history. 
As a result, we gathered 4,150 suspicious tokens, but only 2,723 tokens have Solidity source code available on Etherscan~\cite{Etherscan}. We then manually inspected every selected token contract, which was one of the most time-consuming task in our investigation. 
As a result, we found \textbf{1,859} Trapdoor tokens from the initial suspicious list. These tokens are used as samples to investigates the techniques used by Trapdoor scammers (Section~\ref{subsec:Trapdoor_classification}). 


\subsection{Trapdoor Techniques Classification}
\label{subsec:Trapdoor_classification}
\hg{We classify all malicious techniques found in our sample dataset of 1,859 Trapdoor tokens (Section~\ref{subsec:trapdoor_samples}) into five categories below. Note that multiple Trapdoor techniques can be employed in a single token contract, and it suffices to trigger just one of them to conduct a scam. When inspecting the contract source codes, we also examined the transaction histories of the token creators to confirm if the creators/scammers followed through with their scams using the embedded techniques.} 

\begin{figure}[ht]
\centering
\includegraphics[width=0.48\textwidth]{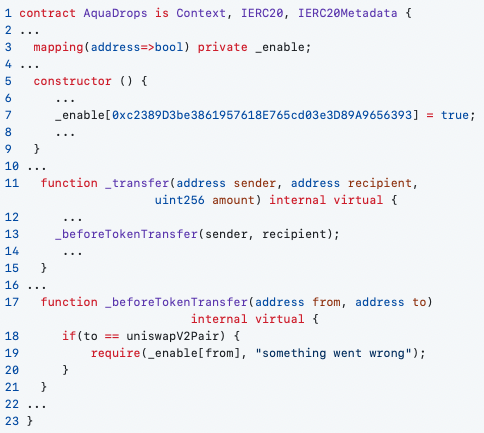}
\caption{Exchange permission example: \textit{AquaDrops}~\cite{AquaDrop}.}
\label{fig:aquadrops}
\end{figure}

\textbf{Exchange Permission.} The token contracts in this group often verify some conditions before allowing the investors to sell. 
For this purpose, either a \textit{black list} or a \textit{white list} is often used.
While a blacklist contains all 
addresses that are forbidden from exchanging tokens, a whitelist stores addresses that have permission to trade in some circumstances. Although these two lists were originally designed to deter transactions from auto-trading programs (sniper bots) or from angel investors in the early stages (prior to DEXs listing), they can also be used 
to restrict selling from investors, while still creating a false sense 
of safety.  To illustrate this technique, let's consider the contract of a Trapdoor token called \texttt{AquaDrops} (see Fig.~\ref{fig:aquadrops}). At first glance, the contract code looks quite normal. An internal function \texttt{\_transfer()} is called by \texttt{transfer} and \texttt{transferFrom()} to support token transferring between two arbitrary addresses. This function then calls the 
function \texttt{\_beforeTokenTransfer()} before starting the transfer (line 13). 
It turns out that this function 
performs a special address check only when a seller sends/sells tokens to the liquidity pool \texttt{uniswapV2Pair}, instead of checking both buyers and sellers (line 18). More specifically, it uses an assert function \texttt{require()} 
to ensure that the seller address is in the eligible list \texttt{\_enable} (line 19). 
If not, an error will be raised, and the sell transaction will be terminated. However, \texttt{\_enable} isn't updated anywhere except 
in the contract's constructor where the address of the creator itself is added to this list (line 7). Thus, the contract creator (scammer) is the only address who can sell a Trapdoor token to the pool to earn the high-value token.

\begin{figure}[ht]
\centering
\includegraphics[width=0.48\textwidth]{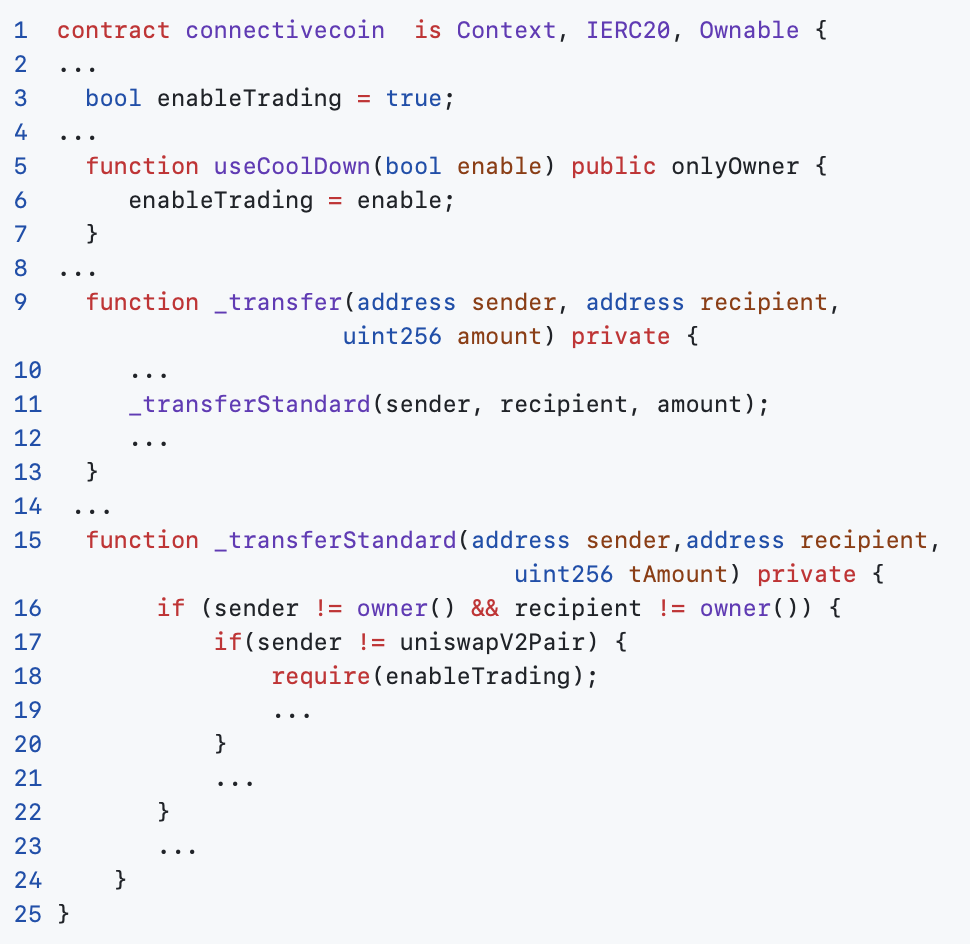}
\caption{Exchange suspension example: \textit{Connective}~\cite{connective}.}
\label{fig:connective}
\end{figure}

\vspace{-5pt}
\textbf{Exchange Suspension.} Unlike the first group, scam tokens in this group employ a suspension mechanism to stop all tradings, 
e.g. by using a boolean variable as a \textit{switch}. The switch was originally turned ``\textit{off}'' so that investors can exchange tokens back and forth. At a particular point, the creator can turn the switch ``\textit{on}'' to stop all tradings except to/from himself. 
An example  
is the \texttt{Connective} token (see Fig.~\ref{fig:connective}). 
At the beginning, the boolean variable \texttt{enableTrading} is 
set to be ``true'' at line 3 of the contract. This variable's value is then asserted inside the function \texttt{\_transferStandard()} (see lines 16-18), which is called by the basic function \texttt{\_transfer()} at line 11. This assertion is applied for every token transfer that is not from an exchange pool \texttt{uniswapV2Pair} or from/to a token creator. In other words, this restriction applies to all sellers except scammers. Initially, the value of \texttt{enableTrading} is \textit{true} so that all investors can freely buy and sell the token. 
However, the value of this variable 
can be updated by using a backdoor function \texttt{useCoolDown()} (line 6) accessible to the creator only. In fact, the scammer did change the value of \texttt{enableTrading} to \textit{false}
(in the transaction \textit{0x8a255e}, effectively preventing everyone except himself from selling the scam token. 

\begin{figure}[ht]
\centering
\includegraphics[width=0.45\textwidth]{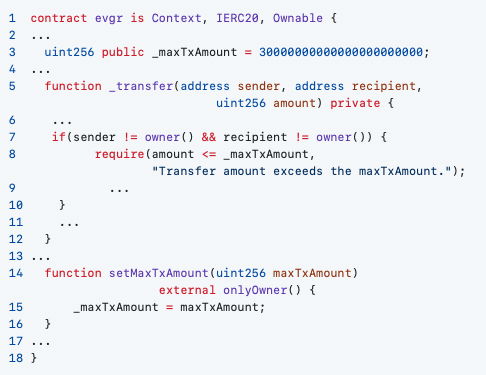}
\caption{Amount limit example: \textit{EVGR}~\cite{evgr}.}
\label{fig:evgr}
\end{figure}

\textbf{Amount Limit.} Instead of checking the exchange permission of addresses or suspending all the tradings, 
token creators/scammers can also employ an exchange limit (integer number) to restrict the transfer amount in each transaction from investors, which can be modified by a backdoor function. 
At the beginning, the limit is set to a large number, letting investors buy the scam token without any restriction. Subsequently, the scammer updates this limit to a very small number (e.g., 0 or 1), effectively preventing selling. For example, the creator of the token \texttt{EVGR}~\cite{evgr} (see Fig.~\ref{fig:evgr}) uses the variable \texttt{\_maxTxAmount} as an exchange limit. This variable was initially set to a very large number (line 3), letting investors buy and sell this scam token without any limitation. Afterward, the scammer updated this limit to 1 
(in the transaction \textit{0xb5a6b0})
using the backdoor function \texttt{setMaxTxAmount()} (lines 14-16). As the result, 
no one except the creator (see line 7) can exchange the scam token due to the assertion at line 8 in the function \texttt{\_transfer()}. 

\begin{figure}[ht]
\centering
\includegraphics[width=0.48\textwidth]{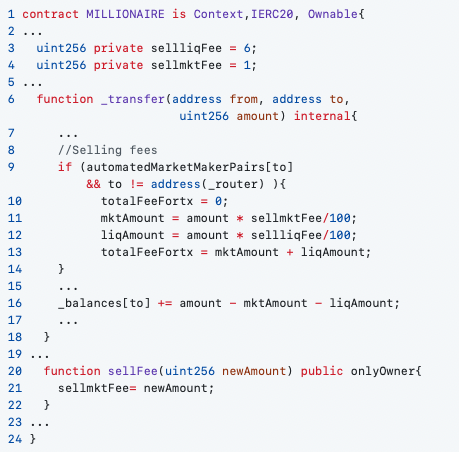}
\caption{Fee manipulation example: \textit{Dollar Millionaire}~\cite{dollarmillionaire}.}
\label{fig:dollar}
\end{figure}

\textbf{Fee Manipulation.} Apart from applying a limit on the sale amount, a scammer can also use a trading fee (integer number) to drain the expected transfer amount. Specifically, tokens in the trading fee manipulation group will charge investors an extremely high trading fee, e.g. by burning a large amount of Trapdoor token every time they buy or sell. As a consequence, after a buy-sell cycle, investors will lose almost all, e.g. 99\%, of the investment. The burning percentage/trading fee ratio can be either defined explicitly in the contract or updated later by the token creator. 
For example, the contract of \texttt{88 Dollar Millionaire} (see Fig.~\ref{fig:dollar}) started with a low fee but raised it later to steal all the investment funds. 
In the contract, the selling fee is the sum of \texttt{sellmktFee} and \texttt{sellliqFee} (line 13). At the beginning, the value of \texttt{sellmktFee} and \texttt{sellmktFee} were 1\% and 6\% respectively (lines 2-3), which are reasonable. However, \texttt{sellmktFee} was later updated to 99\% by the scammer 
(in the transaction \textit{0xb61b83})
using the backdoor function \texttt{sellFee()} at line 20, increasing the fee to 
over 100\%. As a consequence, the investors won't receive any payment back.

\begin{figure}[ht]
\centering
\includegraphics[width=0.48\textwidth]{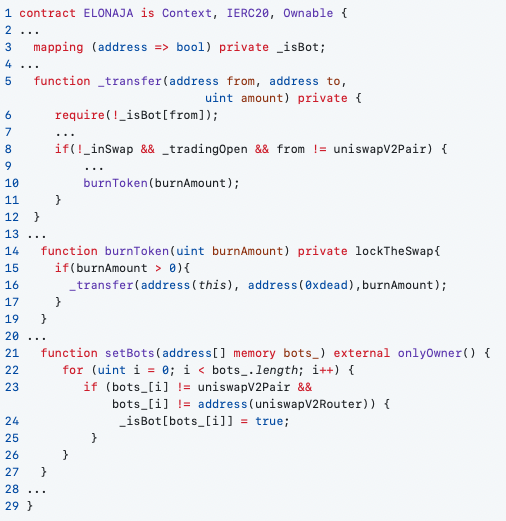}
\caption{Invalid callback example: \textit{ELONAJA}~\cite{ELONAJA}.}
\label{fig:elonaja}
\end{figure}

\textbf{Invalid Callback.} Tokens in this group work as follows. Every time investors call the transfer function to sell a token, a special function is used to call back the transfer function with \textit{invalid} inputs that cause the transaction to fail and roll back. An example of this sophisticated maneuver is \texttt{ELONAJA}~\cite{ELONAJA} (Fig.~\ref{fig:elonaja}). 
The creator of this token first defined an eligible blacklist and named it \texttt{\_isBot} (line 3) to fool investors into thinking that the blacklist was for preventing automated trading bots. This list is used to check if an investor is on the blacklist before doing a token transfer using the assertion function \texttt{require()} (line 6). The list \texttt{\_isBot} is updated by using the function \texttt{set- Bots()} accessible to the token creator only (lines 21-27). The transaction history of the token also indicated that some bot addresses had been added to this list via the transaction 
\textit{0xa64cb3}. At first glance, everything looks normal. However, after checking 
the bot list in the \textit{0xa64cb3} transaction carefully, we found that 
the \texttt{ELONAJA} token address itself was also added to the list (the $8^{th}$ address in the list). 
When a token transfer starts, the address assertion at line 6 is always \rev{satisfied} 
because \rev{the investor address} 
is not in the \texttt{\_isBot} list. 
However, if a transfer is a token sell (line 8), the function \texttt{burnToken()} will be triggered (line 10), which will call the function \texttt{\_transfer()} with the input \texttt{from} set to be \texttt{ELONAJA} token's address, causing an exception raised from the assertion at line 6 in the second call. Thus, token sells are always failed, although seller addresses are not in the blacklist \texttt{\_isBot}.

\vspace{-10pt}
 
\subsection{Trap Placement and Concealment}
\label{subsec:Trapdoor_technique}

We explore in this subsection several important aspects of Trapdoor tokens including where the traps/malicious codes are located (trap location) and how they are concealed (trap concealment) in the source codes.

\textbf{Trap placement.} 
In a Trapdoor token contract, malicious codes can be placed anywhere as long as they are called from the transfer functions. More complex Trapdoor tokens may place their traps in a location other than the transfer functions, and finding them 
in thousands of lines of code is like finding a needle in a haystack. According to our analysis, there are three different locations where a trap is often placed:
\begin{itemize}
    \item \textbf{Modifier:} A \texttt{modifier} is a special type of function whose mission is to perform an additional task before a function is executed (e.g., to validate function inputs). To use a \texttt{modifier}, a developer must attach it to a function by placing its signature next to the function definition, which is often neglected by users while scanning a contract. Hence, a scammer can put 
    malicious codes in a \texttt{modifier} and attach this \texttt{modifier} to the transfer functions, thereby making the trap more hidden. 
    For example, in \texttt{YearnLending.Finance}~\cite{YLF}, a \texttt{onlyPermitted} modifier is attached in the middle of the \texttt{transfer()} function signature, while its definition is in another contract.
    \item \textbf{Function:} Similar to the idea of using a \texttt{modifier}, a scammer can place the malicious codes 
    anywhere and then call them from the transfer functions indirectly across multiple code layers to hide their association with the transfer functions. This trick 
    makes the code inspectors lose track while analysing the source code.
    For example, in the token \texttt{The Reckoning Force}~\cite{TRF}, the transfer fee is calculated from multiple sources and from different nested functions, making it rather challenging to know where the fee comes from and how it is calculated.
    \item \textbf{Contract:} In an even more sophisticated scenario, scammers can place their malicious codes 
    in \textit{another} contract. This will require more advanced detection techniques because inspecting the token's source code alone won't reveal anything malicious. 
    For example, the source code of \texttt{Elongate Deluxe}~\cite{ElongD} calls functions in another contract \texttt{botProtection}, which is where their malicious codes actually are. 
    Those functions' names are encoded in the hex format, 
    making it difficult to tell what they are 
    from the calls alone. In addition, the source code of the external contract \texttt{botProtection} is \textit{unavailable}, which means that contract inspection is impossible. 
    In such cases, we can resort to the transaction histories to find the address of \texttt{botProtection}, and then use EVM decompiler tools (e.g. Panoramix~\cite{Panoramix}) to decompile the contract bytecode to obtain a readable source code.
\end{itemize}

\textbf{Trap concealing.} 
The scammers also use a plethora of tricks and deception techniques to fool the investors and hide the fact that their tokens are scams. 
We discuss below all of those that we discovered when inspecting Trapdoor samples.

\begin{itemize}
\item \textbf{Blank error message:} Using a \texttt{revert} or a \texttt{require} method with a blank message that gives investors no information when they receive an error, making them clueless about the real cause of the transaction failure.
An example of tokens using this technique is \texttt{Takeoff}~\cite{Takeoff}.

\item \textbf{Single-character names:} Using hard-to-see letters such as \texttt{i}, \texttt{t} or \texttt{l} for a switch's name that could be overlooked when reading a contract. Furthermore, searching for a single letter in a contract will yield many unrelated results, making it harder to track the updates of the switch' value. A good example of a scam token using this trick is \texttt{Teen Mutant Turtle}~\cite{TMT}.
\item \textbf{Misleading names:} Scam token creators very often name their malicious variables or functions with misleading names to deceive the investors.
For instance, they name a blacklist as a bot list or an exchange pool address as a zero address, which confuses 
the investors. 
An example of such tokens is \texttt{The Art}~\cite{TheArt}.
\item \textbf{Dummy function:} Token creators sometimes create dum-my functions to show users that the value of a particular variable is updated correctly, and such functions often have standard names such as \texttt{init}. Nevertheless, such function will never be called, and the variable will always have the value the creator wants, serving the purpose of the scam 
(e.g. \texttt{AIRSHIB}~\cite{AIRSHIB}).
\item \textbf{Incomplete renouncement:} Contract ownership renoun-cement, in which the contract creator renounce their ownership of the tokens \hg{by burning their LP tokens}, is a standard way to build trust from investors. However, scammers can actually exploit this action to create \textit{false} trust. For example, before giving up ownership, the scammer can secretly add another of his addresses, which still allows him to manipulate the contract without the ownership.
Therefore, although investors can see that the contract creator has relinquished the ownership, the token is still secretly under his control (e.g. \texttt{DYDZ}~\cite{DYDZ}). 
\item \textbf{Numerical exception:}
Scam tokens using this technique also tend to cause errors to terminate transactions on pre-set conditions. However, instead of causing an error actively using \texttt{require}, a \textit{numerical exception} occurs due to the invalid values intentionally updated by the scammer. This trick may make the investors think that some honest numerical bug was the cause, not that the token is a scam.
For instance, the contract of \texttt{CPP4U}~\cite{cpp4u} uses a variable \texttt{totalSellTax} to record the trading fee percentage, which was originally set to 0. 
However, the token creator secretly increased this value to 1000\%, resulting in the \texttt{fee} ten times higher than the value \texttt{\_amount} of the transaction.
The subtraction of \texttt{fee} from \texttt{\_amount}, therefore, will be a negative number, 
which is then assigned to \texttt{rAmount}, an unsigned integer.
This assignment causes the exception ``\texttt{Integer overflow}'', making it look like a programming bug.
\end{itemize}

\section{\hg{A Reliable Trapdoor Detection Tool and a Large Trapdoor Dataset with Ground Truth}}
\label{sec:Trapdoor_dataset}

\hg{Capitalizing on our previous
Trapdoor token analysis and classification, we first develop \textit{TrapdoorAnalyser}, a novel Trapdoor detection tool that examines and crosschecks both the error-log of a buy-and-sell test and the Trapdoor-indicator list generated by a contract-semantic check, thus providing a more reliable and finer-grained approach compared to existing tools in the industry.
For example, by contrast to well-known tools such as GoPlus~\cite{goplus} and TokenSniffer~\cite{tokensniffer}, TrapdoorAnalyser can output specific variables or functions that prevent the investors from selling. Moreover, thanks to its fine-grained approach, TrapdoorAnalyser can identify advanced Trapdoor techniques that other tools fail to capture (see Section~\ref{subsec:evaluation} for examples).  
Our second contribution in this section is to use TrapdoorAnalyser to build a reliable benchmark dataset of about 30,000 tokens in total \textit{with ground truth} for the Trapdoor detection problem, which 
allows us to train machine-learning models that can deal with scam tokens that have no source codes.} 

\hg{TrapdoorAnalyser consists of three main components described as follows.} (1) The \textit{Buy-and-Sell-Check} component 
verifies if a token and its corresponding exchange pool are currently operating on the chain like a Trapdoor scam (e.g. one can buy the token but cannot sell, or the token charges an extremely high exchange fee). 
It also collects the trading 
outputs, including buy/sell results, messages, and a stack trace-back of all errors that occurred. (2) The \textit{Contract-Semantic-Check} component 
analyses the syntax in the token's contract and collects all 
embedded Trapdoor techniques 
\hg{(e.g. black list, switch)} that meet predefined conditions derived from the characteristics of Trapdoor listed in Section~\ref{subsec:Trapdoor_classification} and Section~\ref{subsec:Trapdoor_technique}. Such techniques are labelled as Trapdoor indicators. 
(3) The \textit{Ground-Truth-Labelling} component labels tokens by 
\hg{matching} the exchange outputs obtained from the Buy-and-Sell Check 
and their Trapdoor indicators from the Contract-Semantic Check. 

\hg{We will describe each of the three components of TrapdoorAnalyser in detail in Sections~\ref{subsec:buy_sell_check},~\ref{subsec:semantic_check}, and~\ref{sec:labelling}. The evaluation of TrapdoorAnalyser is carried out in Section~\ref{subsec:evaluation}, in which we compare our tool directly with  GoPlus~\cite{goplus}.} 


\subsection{\hg{Buy-and-Sell Check}}
\label{subsec:buy_sell_check}

The Buy-and-Sell Check \hg{component of TrapdoorAnalyser}  
was implemented using Brownie \cite{Brownie}, a Python-based development and testing framework for Ethereum smart contracts. 
It is used for determining if a token \hg{can still be traded with a reasonable fee, which includes two parts: } 
\textit{trading result assertion} and \textit{trading fee assertion}. For \textit{trading result assertion}, we check if a token was successfully bought or sold by an investor. We expect that a Trapdoor token can pass a buy test but fail in a sell test, while a non-Trapdoor token can pass both tests. The \textit{trading fee assertion} is only conducted if \hg{both buy and sell were successful}, which 
verifies if the trading fee that the user was charged directly from the transferring amount is affordable. According to popular online audit tools~\cite{goplus,honeypotis,detecthoneypot}, acceptable fee thresholds vary from 10\% to 50\%. Therefore, in our study, we set the threshold for this fee at 30\% of the sending amount. Similarly to the previous assertion, we expect that a Trapdoor token contract will charge a very small fee on buying transactions to encourage users to buy this token, while the selling fee should be extremely high to prevent them from retrieving high-value tokens. The token buy/sell test results, buy/sell taxes, and error logs (stack trace-back) are stored and further used in the \textit{ground truth labelling} step. More details about the Buy-and-Sell Check can be found in Procedures~\ref{algo:buy_test} and~\ref{algo:sell_test} below.  

\floatname{algorithm}{Procedure}
\begin{algorithm}[htb]
\caption{\textbf{TokenBuyingTest}$(\textsf{investor}, \textsf{pool}, \textsf{token}, \textsf{acc\_fee})$}
\begin{algorithmic}[1]
    \STATE $\textsf{liquidity} \gets \textsf{token}.$\textbf{balanceOf}($\textsf{pool}$)
    \STATE $\textsf{last\_balance} \gets \textsf{token}.$\textbf{balanceOf}($\textsf{investor}$)\\
    $\textsf{amount} \gets \min\{\textsf{liquidity}, 1000\}$
    \STATE $\textsf{result} \gets \textsf{token}$.\textbf{transfer}($\textsf{investor}$, $\textsf{amount}$, $\textsf{pool}$)
    \IF[fail to buy $\textsf{amount}$ tokens]{$!\textsf{result}$}
        \RETURN $\mathsf{false}$
    \ENDIF
    \STATE $\textsf{expected\_amount} \gets \textsf{amount} \times (\mathsf{1}- \textsf{acc\_fee})$
    \STATE $\textsf{current\_balance} \gets \textsf{token}$.\textbf{balanceOf}($\textsf{investor}$)
    \STATE $\textsf{received\_amount} \gets \textsf{current\_balance} - \textsf{last\_balance}$
    \IF[high fee]{$\textsf{received\_amount}\hspace{-2pt}<\hspace{-2pt}\textsf{expected\_amount}$}
        \RETURN $\mathsf{false}$
    \ENDIF
    \RETURN $\mathsf{true}$
\end{algorithmic}
\label{algo:buy_test}
\end{algorithm}
\vspace{-5pt}

\floatname{algorithm}{Procedure}
\begin{algorithm}[htb]
\caption{\textbf{TokenSellingTest}$(\textsf{investor}, \textsf{pool}, \textsf{token}, \textsf{acc\_fee})$}
\begin{algorithmic}[1]
    \STATE $\textsf{balance} \gets \textsf{token}.$\textbf{balanceOf}($\textsf{investor}$)
    \STATE $\textsf{last\_liquidity} \gets \textsf{token}.$\textbf{balanceOf}($pool$)
    \STATE $\textsf{amount} \gets \textsf{balance} \times \mathsf{0.5}$
    \STATE $\textsf{result} \gets \textsf{token}$.\textbf{transfer}($\textsf{pool}$, $\textsf{amount}$, $\textsf{investor}$)
    \IF[fail to sell $\textsf{amount}$ tokens]{$!\textsf{result}$}
        \RETURN $\mathsf{false}$
    \ENDIF
    \STATE $\textsf{expected\_amount} \gets \textsf{amount} \times (\mathsf{1}- \textsf{acc\_fee})$
    \STATE $\textsf{current\_liquidity}\gets \textsf{token}.$\textbf{balanceOf}($\textsf{pool}$);
    \STATE $\textsf{received\_amount} \gets \textsf{current\_liquidity} - \textsf{last\_liquidity}$;
    \IF[high fee]{$\textsf{received\_amount} \hspace{-2pt}<\hspace{-2pt} \textsf{expected\_amount}$}
        \RETURN $\mathsf{false}$
    \ENDIF
    \RETURN $\mathsf{true}$
\end{algorithmic}
\label{algo:sell_test}
\end{algorithm}

\subsection{Contract-Semantic Check}
\label{subsec:semantic_check}

The Contract-Semantic Check \hg{component of TrapdoorAnalyser} is made up of two sub-components: \textit{Contract Analysis} and \textit{\duy{Trapdoor-indicator} Detection}. 
Contract Analysis is built on top of \textit{Slither}~\cite{feist2019slither}, the Solidity static analysis framework written in Python that allows us to write custom analysis via 
their public APIs. 
It 
first constructs the \textit{Abstract Syntax Tree (AST)}~\cite{ast} from the contract's Solidity source code. The \textit{AST} is a tree-like data structure where each node in a tree represents a programming element such as a variable, an operand, or a function, and an edge in a tree indicates the dependence between the corresponding two nodes/elements. We analyse a token's contract by traversing its 
AST and gathering useful information, 
including a list of state variables and their dependants $S$, a list of transfer functions and their related \rev{functions/modifiers $F$}, a list of transfer functions' inputs and their dependants $I$, a list of end nodes $N$, a list of backdoor functions $B$, and a list of external contract $C$. 
Here, the dependants of a variable $X$ are the variables whose values are influenced by the value of $X$.  More details are provided below. 

\begin{itemize}
    \item \textbf{State variables ($S$)} are variables created globally \duy{at} a contract level. These variables can be called in any function within a contract. A list of state variables and their dependants are stored in $S = \{s_a, s_i, s_b, s_c\}$ where $s_a$ denotes address-type variables or lists of addresses, $s_i$ specifies integers, $s_b$ are booleans, and $s_c$ are constants.

    \item \textbf{Transfer functions ($F$)} includes two functions \texttt{transfer}- \texttt{From()} and \texttt{transfer()} which must be implemented from the interface \hg{ERC-20}. The set of function $F$ includes not only these two functions but also functions/modifiers called from them.
    
    \item \textbf{Transfer inputs ($I$)} consist of three different variables $I = \{i_{s}, i_{r}, i_{a}\}$, where $i_{s}$ is a sender address,  $i_{r}$ is a receiver address, and $i_{a}$ denotes a transfer amount.

    \item \textbf{End nodes ($N$)} will \rev{stop the contract execution when reached}. End nodes consist of assertion nodes (such as \texttt{require} or \texttt{assert}) and exit nodes (e.g. \texttt{revert} or \texttt{return}). We only collect end nodes that are used in $F$. 

    \item \textbf{Backdoor functions ($B$)} are functions that can be executed by a specific address (e.g., a token owner). We identify a backdoor function by checking if it verifies the permission of a caller by \duy{comparing with} a predefined address \duy{or checking if a caller's address is available in a list of addresses}.

    \item \textbf{External contract calls ($C$)} are the function calls to functions in other contracts that are not defined within a token's source code. \duy{We obtain these calls by extracting all \texttt{low\_level\_call}
    \footnote{https://solidity-by-example.org/calling-contract/} node.}

\end{itemize}

After gathering all of the above information, the \textit{\duy{Trapdoor-indicator} Detection} 
will be run to collect all \duy{Trapdoor indicators} that may be used by a scammer to trap the funds of users. The check consists of five separate sub-checks, each of which 
is used to detect \duy{Trapdoor indicators} for a particular \duy{group} listed in Section~\ref{subsec:Trapdoor_classification}. The implementation of each sub-check is described below. 

\begin{itemize}
    \item \textbf{Exchange Permission.} We detect \duy{Trapdoor indicators} used for conducting exchange-permission Trapdoor by iterating over all variables $s_a$ in $S$. An $s_a$ is reported as a \duy{Trapdoor indicator} if it \duy{is used to check the permission of a sender address} $i_s$ or \duy{a receiver address} $i_r$ in \duy{any end node} $n \in N$, and moreover, this $s_a$  can be modified by \duy{a backdoor function} in $B$ or \duy{an external contract call} in $C$.
    \item \textbf{Exchange Suspension.} 
    We first collect all \duy{boolean variables} $s_b$ contained in $S$. A variable is then reported as a Trapdoor indicator 
    if it is used in any end node in $N$, and moreover, \duy{its value can be modified by a backdoor function in $B$ or an external contract call in $C$}. 
    \item \textbf{Amount Limit.} 
    We examine all numeric variables $s_i\in S$ and report any $s_i$ as a \duy{Trapdoor indicator if it is used to compare with a transfer amount} $i_a$ in any end node $n$, and moreover, the value of $s_i$ 
    \hg{can be modified} using a backdoor function in $B$ or an external contract call in $C$.
    \item \textbf{Fee Manipulation.} We report a numeric variable $s_i\in S$ or constant variable $s_c\in S$ as a Trapdoor indicator 
    if it is used to \duy{calculate and update the value of} \rev{a transfer input} 
    $i_a$, and moreover, it can be modified by using a backdoor function in $B$ or a contract call in $C$.
    \item \textbf{Invalid Callback.} Detecting an invalid callback is \duy{more challenging}. \hg{We need to 
    examine all functions in $F$ and build a \textit{directed graph} $G=(V,E)$ where $V$ represents the set of all functions and $E$ represents all function calls}. For example, an edge $v_1 \rightarrow v_2$ means a function $v_2$ is called in $v_1$'s body. A function $v_k$ \duy{is marked as a Trapdoor indicator}, if it satisfies the following conditions: (1) \hg{$G$ has a cycle containing $v_k$}, (2) one of $v_k$'s inputs is compared with $s_i$ (or $s_a$) in any end node in $N$, and (3) the item $s_i$ (or $s_a$) is managed by \hg{the token creator} 
    using a backdoor function in $B$ or an external contract call in $C$. In this way, a scammer can manipulate the value of the input when calling $v_k$ so that the comparison of the input of $v_k$ and $s_i$ (or $s_a$) at the end node is invalid, disrupting the token transfer process.
\end{itemize}

\subsection{Ground Truth Labelling}
\label{sec:labelling}

\hg{We discuss in this section Ground-Truth Labelling, the third component of TrapdoorAnalyser, and explain how to build the first Tradoor dataset on UniswapV2 using this tool. We also identify a drawback in existing datasets of related scams.}

\textbf{Non-Trapdoor Tokens.} 
\rev{Most existing} studies, e.g.~\cite{Xia:2021,mazorra2022not,nguyen2023rug}, collected non-malicious tokens 
based on their popularity in the cryptocurrency market. 
\rev{Such} tokens \rev{received} 
high ranks on some reputable websites (e.g., Coinmarketcap~\cite{coinmarketcap}) and \rev{had} been purchased by many investors. 
\rev{Although it seems quite plausible at first glance, this approach has serious flaws.}  

\textit{First}, popular tokens often have very active transaction histories with frequent trading activities, whereas scam tokens, \rev{similar to those unpopular ones, usually have 
very short and sparse transaction \rev{histories}~\cite{Xia:2021,mazorra2022not,cernera2023token}. 
Hence, using popular tokens as non-malicious samples inadvertently creates a low-quality dataset and inaccurate machine learning models, which can discriminate well a malicious token from a popular but not from an unpopular token.}

\textit{Second}, a popular token might still contain bugs or codes that may be exploited to turn it into a Trapdoor.
For example, the Lido DAO Token (LDO)~\cite{lido} was ranked $57^{th}$ on the Coinmarketcap website at the time of this study. This token had been listed since May 2021 and traded by many investors. At first sight, it looks perfectly safe 
from the market capitalisation and popularity perspectives. However, the token contains the \duy{Trapdoor indicator} \texttt{transfersEnabled} (line 168 in~\cite{lido_eth}) that could be used for suspending the token exchange by its creator via the creator-only function \texttt{enableTransfers} (line 448 in~\cite{lido_eth}). 
Although the creator might have created the token without any malicious intent, there is still a risk that it can be turned into a Trapdoor token at some point. 

On the contrary, TrapdoorAnalyser 
doesn't label a token based on its popularity. 
In fact, it labels
non-Trapdoor tokens based on the results of the checks described in Sections~\ref{subsec:buy_sell_check} and~\ref{subsec:semantic_check}, which is more systematic and reliable. 
More specifically, if a token passes the \textit{Buy-and-Sell Check} and no \duy{Trapdoor indicators} are found in its contract by the \textit{Contract-Semantic Check}, then it 
will be marked as ``non-Trapdoor''. 
Applying TrapdoorAnalyser to the token dataset built in Section~\ref{subsec:tokens_pools_datasets}, we obtained \textbf{18,548} non-Trapdoor tokens. \hg{Note that our method doesn't capture every non-Trapdoor token. For example, those that did not pass the Buy-and-Sell Check for some unknown reasons will not be labelled. However, we chose to err on the conservative side to ensure that the list of non-Trapdoor tokens we obtained is correct.}

\begin{figure}[ht]
\centering
\includegraphics[width=0.35\textwidth]{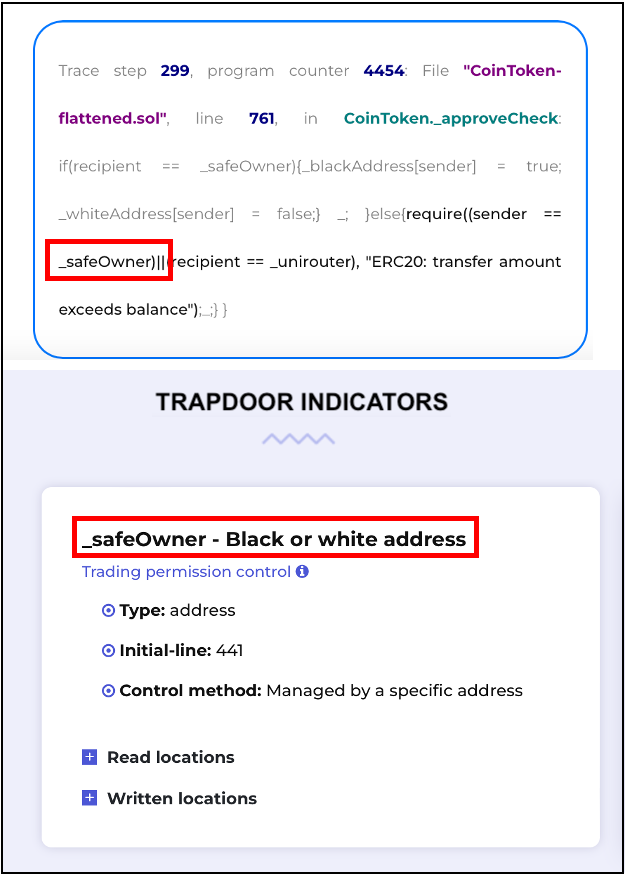}
\caption{Trapdoor labelling example.}
\label{fig:trapdoor_labelling}
\vspace{-5pt}
\end{figure}

\textbf{Trapdoor Tokens.} 
A token is labelled by TrapdoorAnalyser as a Trapdoor if it meets the following two criteria simultaneously: (1) a token fails the sell test, and (2) the failure is caused by one of the \duy{Trapdoor indicators} found by the \textit{Contract-Semantic Check}. 
Note that a scam token may pass or fail the buy test. 
For example, the buy test would fail if the token creator has suspended the entire token trading (buy/sell) or applied a considerable trading fee for both buy and sell transactions, which makes the buyer lose their funds instantly after buying, resulting in a buy failure. 
If both buy and sell tests fail, we extracted their corresponding error logs and checked if any \duy{Trapdoor indicator} obtained from the Contract-Semantic Check is listed in the logs. If there is, we will label a token as ``Trapdoor''. An example of this process 
is shown in Fig.~\ref{fig:trapdoor_labelling}, \hg{in which the item \texttt{\_safeOwner} in the error log of the Buy-and-Sell Check matches a Trapdoor indicator (black/white list) found by the Contract-Semantic Check}. Eventually, we obtained \textbf{11,943} Trapdoor tokens, including those Trapdoor samples found in Section~\ref{subsec:trapdoor_samples}. 
For tokens that violate the labelling conditions of ``Trapdoor''/``non-Trapdoor'', we labelled them as ``unknown''. 

\subsection{Method Evaluation}
\label{subsec:evaluation}

In this section, we evaluate the effectiveness and performance of TrapdoorAnalyser in comparison with the state-of-the-art tool from industry - GoPlus~\cite{goplus}. In this section, we only give the comparison with GoPlus, but not other popular tools such as Token Sniffer~\cite{tokensniffer} or ``honeypot.is''~\cite{honeypotis} due to some reasons. First, a ``honeypot.is'' verifies a token primarily based on buy-and-sell check. Therefore, if this tool cannot buy a token due to some reason such as empty liquidity or a scammer suspended a token trading, it cannot verify a given token. Taking the FXF token~\cite{FXF} as a good example, ``honeypot.is'' returns ``UNKNOWN'' when it cannot do a buy check. Another drawback of this tool is that it can detect Trapdoor tokens only after these scams have occurred. This tool also admits that it is not a foolproof method because a token is considered non-scam at the moment does not mean it will not change in the future. Similarly, Token Sniffer uses a detecting API from ``honeypot.is'' for checking Trapdoor so it inherits all shortcomings of ``honeypot.is''. Although this tool further analyse a token profile (e.g, holders, liquidity) via transaction history, these analyses do not provide information for detecting Trapdoor. Different from these tools, GoPlus provides APIs~\cite{goplus_api} for checking the security risks~\cite{goplus_security_items} of a token by analysing its smart contracts. Some of the risks can represent Trapdoor traits. Therefore, for the fair comparison between our tool and GoPlus, we map four of the Trapdoor groups in our study, including Exchange Permission (EP), Exchange Suspension (ES), Amount Limit (AL), and Fee Manipulation (FM), 
to seven equivalent GoPlus securities items (see Table~\ref{tab:goplus_mapping}). 
\hg{We first compare the performance of both tools on our large ground-truth dataset, and then on a sampling dataset of 290 unknown tokens (outside our dataset but manually inspected and labelled by us for this comparison).} 


\begin{table}[htb!]
\centering
\caption{The mapping between TrapdoorAnalyser's groups and GoPlus' security items, e.g. TrapdoorAnalyser's EP is equivalent to (is\_blacklisted OR is\_whitelisted) in GoPlus. A dash `-' means there are no equivalent items.}
\label{tab:goplus_mapping}
\begin{tabular}{|c|c|l|}
\hline
\textbf{Types}  & \textbf{Logical}     & \textbf{GoPlus items}   \\ \hline
\multirow{2}{*}{\textbf{EP}} & \multirow{2}{*}{OR}& is\_blacklisted        \\ \cline{3-3} 
                             &   & is\_whitelisted         \\ \hline
\multirow{2}{*}{\textbf{ES}}&\multirow{2}{*}{OR} & transfer\_pausable      \\ \cline{3-3} 
                  &  & trading\_cooldown \\ \hline
\multirow{2}{*}{\textbf{AL}}&\multirow{2}{*}{AND} & is\_anti\_whale        \\ \cline{3-3} 
                   & & anti\_whale\_modifiable \\ \hline
\textbf{FM}        &     \hg{N/A}     & tax\_modifiable         \\ \hline
\textbf{IC}         &    -     & -                       \\ \hline
\end{tabular}
\vspace{-5pt}
\end{table}


\textbf{Trapdoor Detection Effectiveness 
} 
We first compare the effectiveness of the Contract-Semantic Check component of our TrapdoorAnalyser and GoPlus in identifying known/unknown Trapdoor/ non-Trapdoor tokens. 
To this end, 
we run GoPlus on our ground-truth dataset, 
the outcome of which is reported in Table~\ref{tab:goplus_groundtruth}. 
\hg{Note that as long as GoPlus identifies the presence of one risky item (in their language), we count the token as a Trapdoor detected by GoPlus}. 
Notably, while both tools can identify all \textit{non-Trapdoor} tokens, 
GoPlus can only successfully detect 60\% of 
\textit{Trapdoor} tokens in our dataset. 
According to our observation of GoPlus's detection results, GoPlus's audit mission is to report all items in a token contract that may pose a risk for users by matching known patterns. These patterns become obsolete due to the evolution and diversity of code structures and obfuscation techniques. 
On the contrary, our semantic-based method mainly focuses on collecting and validating items that directly impact the transfer process of a token regardless of the code structure. Therefore, our tool can detect more hidden Trapdoor variants than GoPlus. 

For example, in the token \texttt{YearnLending.Finance}~\cite{YLF}, the \texttt{\_permitted} blacklist is used for checking the permission of the sender before conducting a token transfer. This blacklist is defined in another contract and applied to the transfer process through a \texttt{modifier}. This list is also managed by a scammer by using a backdoor function \texttt{givePermission()}. In this case, the blacklist is not defined in the main contract and is not called directly in transfer functions. Therefore, GoPlus cannot distinguish this blacklist's impact on the token transfer process and report this risk. 
Another good example is the token \texttt{AIRSHIB}~\cite{AIRSHIB}. The creator of this token does not use any backdoor function to control a trap. This creator applies the always-fail condition for the token sell checks in transfer functions and bypasses it by using some white addresses. These addresses are overlooked by GoPlus because this tool cannot find any malicious functions to update these addresses and it also cannot check if these addresses are used to interrupt the token transfer process. Finally, GoPlus is also unable to detect IC cases because a calling ``cycle'' can only be detected by checking functions execution flow, which is not the focus of this tool. 

\begin{table}[htb!]
\centering
\caption{Comparison of GoPlus and TrapdoorAnalyser on the ground-truth dataset.}
\label{tab:goplus_groundtruth}
\begin{tabular}{|l|c|c|}
\hline
                    & \textbf{TrapdoorAnalyser} & \textbf{GoPlus} \\ \hline
\textbf{Non-Trapdoor}        & 18,548                 & 18,548           \\ \hline
\textbf{Trapdoor}            & 11,943                 & 7,172           \\ \hline
\end{tabular}
\end{table}

\textbf{Injected Techniques Detection Accuracy.}
\hg{Next, we check if TrapdoorAnalyser has a better detection accuracy than GoPlus on a dataset of previously unlabelled tokens. To this end, we run GoPlus and our tool on a sampling dataset. For a fair comparison, this dataset consists of 290 tokens sampled from the ``unknown'' set (see Section~\ref{sec:labelling}) whose addresses 
ending with (a randomly chosen pair) `df'. 
We then inspected these token contracts and manually labelled them. Note that these tokens satisfy neither ``Trapdoor'' nor ``non-Trapdoor'' conditions in TrapdoorAnalyser, hence not included in our full-size dataset.} The rates of true positive (TP), false positive (FP) and false negative (FN) of both tools are reported in Table~\ref{tab:goplus_sampling}. 


\begin{table}[htb!]
\centering
\caption{A comparison of detection performance between TrapdoorAnalyser and GoPlus on a sampling dataset (``unknown'' tokens ending with `df').}
\label{tab:goplus_sampling}
\begin{tabular}{|c|c|ccc|ccc|}
\hline
\multirow{2}{*}{\textbf{\begin{tabular}[c]{@{}c@{}}Trapdoor \\ indicator\end{tabular}}} & \multirow{2}{*}{\textbf{Total}} & \multicolumn{3}{c|}{\textbf{TrapdoorAnalyser}}                                    & \multicolumn{3}{c|}{\textbf{GoPlus}}                                              \\ \cline{3-8} 
                                                                                        &                                 & \multicolumn{1}{c|}{\textbf{TP}} & \multicolumn{1}{c|}{\textbf{FP}} & \textbf{FN} & \multicolumn{1}{c|}{\textbf{TP}} & \multicolumn{1}{c|}{\textbf{FP}} & \textbf{FN} \\ \hline
\textbf{EP}                                                                             & 211                             & \multicolumn{1}{c|}{211}         & \multicolumn{1}{c|}{0}           & 0           & \multicolumn{1}{c|}{188}         & \multicolumn{1}{c|}{0}           & 23          \\ \hline
\textbf{ES}                                                                             & 182                             & \multicolumn{1}{c|}{178}         & \multicolumn{1}{c|}{0}           & 4           & \multicolumn{1}{c|}{107}         & \multicolumn{1}{c|}{0}           & 71          \\ \hline
\textbf{AL}                                                                             & 205                             & \multicolumn{1}{c|}{204}         & \multicolumn{1}{c|}{0}           & 1           & \multicolumn{1}{c|}{150}         & \multicolumn{1}{c|}{0}           & 55          \\ \hline
\textbf{FM}                                                                             & 137                             & \multicolumn{1}{c|}{137}         & \multicolumn{1}{c|}{0}           & 0           & \multicolumn{1}{c|}{132}         & \multicolumn{1}{c|}{0}           & 5          \\ \hline
\textbf{IC}                                                                             & 6                               & \multicolumn{1}{c|}{6}           & \multicolumn{1}{c|}{0}           & 0           & \multicolumn{1}{c|}{-}           & \multicolumn{1}{c|}{-}           & -           \\ \hline
\end{tabular}
\end{table}

On the sampling dataset, TrapdoorAnalyser can detect indicators of all EP, FM, and IC, 99.5\% of AL and 97.8\% of ES Trapdoor tokens. 
The detection results of GoPlus are 89.0\%, 58.0\%, 73.1\%, and 96.3\% of EP, ES, AL, and FM tokens, respectively. 
Moreover, our tool can detect all Trapdoor risks that GoPlus can catch except one ES technique, while GoPlus fails to report some variants of Trapdoor. Particularly, it misses 23 tokens that contain EP indicators. As a concrete example, 
the token \texttt{ySell}~\cite{ySell} managed to bypass GoPlus' check by naming its whitelist with a misleading name \texttt{allow}, which is very similar to the \texttt{allowed} list used for storing the wallet spending limit approved by a token holder. 

GoPlus also fails to report 71 ES Trapdoor tokens, including \texttt{KAMARERU}~\cite{KAMARERU}, in which the creator 
also 
names an exchange pool address by a misleading name \texttt{router} to conceal the trap from investors that the receiver address will be compared with the pool address (a sell transfer check) and jump into the malicious code block if these addresses match 
Moreover, the creator also conceals a trap by naming a switch \texttt{balances1} and a whitelist \texttt{\_balances1} similar to token balances storage (\texttt{\_balances}). In contrast, our tool can detect these cases because it examines 
a variable based on its usage in the token transfer flow and not on its name. Our tool only fails to identify one variant of ES (4 FNs) because in these tokens, a switch is used indirectly with the assertion function nested in many \texttt{if} conditions. TrapdoorAnalyser doesn't associate an indicator with end nodes that are located too far away to avoid the case of false positives. 

For AL, GoPlus has 55 FN cases, while we have only one FN case. An example of FN cases of GoPlus is \texttt{DOGEONE}~\cite{DOGEONE} token. In this token, a scammer defines an object (\texttt{struct}) \texttt{Limits} for storing buying limit and wallet amount limit instead of using two different primitive-type variables, making it hard for GoPlus 
to analyse and verify the data flow of this indicator. For our FN case~\cite{THECLASSICMEME}, the amount limit calculation in this token is highly similar to the transfer fee calculation. Therefore, this indicator is mislabelled as FM. 

For FM detection, the performances of our tool and GoPlus are almost the same. There are only 5 cases of FM that are not reported by GoPlus because these cases contain the complex tax calculation flow. Take \texttt{SHIBAPPE}~\cite{SHIBAPPE} token as an example. The sell and buy taxes of this token are calculated and updated in many places inside and outside the transfer functions. Moreover, the final sell tax is aggregated and split multiple times from different taxes, making the execution flow more complex. Hence, GoPlus fails to report these cases because it doesn't have an analysis of execution flow in its detection method. 

\hg{In conclusion, TrapdoorAnalyser 
outperforms GoPlus and 
can identify more embedded Trapdoor techniques by analysing risks that impact the token transfer process.} 

\section{Machine Learning Based Detection}
\label{sec:ML_based_detection}
Although TrapdoorAnalyser 
performs exceptionally well in building the ground truth dataset and identifying Trapdoor tokens, \rev{it only works on} 
tokens that have Solidity source \rev{codes}. According to Fig.~\ref{fig:token_and_exchange_pool}, about 23\% tokens on Uniswap do not \rev{have source codes available.} \duy{Moreover, according to Etherscan, nearly 99\% of contracts on Ethereum do not make their source codes public~\cite{verified_contract}.}
Therefore, relying \rev{entirely} on Solidity source-code analysis will limit the applicability of this approach in general. 

To overcome the aforementioned shortcoming, we adopt the machine learning approach, which has been used widely in detecting different scams on Ethereum effectively, to build a detection tool based on our ground truth dataset. In order to make our tool applicable to all smart contracts, we train our detection model relying on features extracted from two basic and always-available units of blockchain: transactions and operation codes (opcodes). 
We conduct several experiments to evaluate the tool from different perspectives, including comparing the performance of detection models in detecting Trapdoor tokens and identifying 
the best models and the most effective categories of features. Then, from this top-performing model, we analyse the top ten discriminative features the model used to gain a better understanding of how these features contribute to the model performance. Finally, we verify the ability of models to identify Trapdoor techniques embedded 
in a smart contract. 



\subsection{Exchange Features}
According to their illicit behaviour, Trapdoor tokens have several distinct characteristics compared with normal tokens: (1) Most Trapdoor tokens often have a short lifetime; (2) Some tokens can only be bought but not sold, making the number of buyers dominantly larger than the number of sellers; (3) An extreme tax may be applied when investors buy or sell a Trapdoor token, resulting in a large amount of a paired valuable token flowing into an exchange pool but not coming out. Moreover, the main behaviour of a token can be reflected by its exchange history. Therefore, we extract the following features from the event logs of a token and its exchange pools: \textbf{n\_transfers} is the total number of token transfers between two arbitrary addresses, \textbf{n\_token\_transfers} is the total number of transfers from or to the token address and \textbf{r\_token\_transfers} is the ratio of \textbf{n\_token\_transfers}, \textbf{n\_creator\_transfers} (resp. \textbf{r\_creator\_transfers}) is the total number of token transfers from or to token's creator address, \textbf{life\_time} is the token's lifetime - the period from the recreation time of a token to its last transfer transaction, \textbf{n\_swap} (resp. \textbf{n\_sync}, \textbf{n\_burn}, \textbf{n\_mint}) is the number of swap (resp. sync, burn, and mint) events, \textbf{n\_users} is the number of people who exchange a token, \textbf{n\_providers} is the number of liquidity providers, \textbf{n\_buyers} (resp. \textbf{n\_sellers}) is the number of people who exchange a high-value token for a trapdoor token (resp. vice versa), 
 \textbf{sell\_amt} (resp. \textbf{buy\_amt}) is the amount of a token go in (resp. out) a pool, and \textbf{pool\_life\_time} is the pool's lifetime - the period from the recreation time of a pool to its last event.


\subsection{Code Features}
The operation code (opcode) is very useful in analysing the logic of the
smart contract. Opcode has been proven to be quite successful in detecting Ponzi, Honeypot and other scams on blockchain~\cite{chen2018detecting,chen2021sadponzi,torres2019art}. Hence, we expect that features extracted from opcode will also be useful in detecting Trapdoor tokens. The intuition for using opcode features is that they should look quite different between contracts that have different purposes. Moreover, opcode features are not impacted by the token popularity and can be extracted even when the source code (in Solidity) of the smart contracts has been removed. To aggregate opcode features, we collect all contracts' bytecode and disassemble the bytecode into opcodes by using Ethereum's python library \textit{evmdasm}\footnote{https://github.com/ethereum/evmdasm}. Finally, we calculate the occurrence frequency of each unique opcode in each contract (term frequency). Each opcode with its frequency is used as a feature. As a result, \textbf{142} different code features are collected from 30,491 token contracts in our dataset.

\subsection{Experiment Setup}
\label{subsec:experiment}

\textbf{Experimental datasets:}
 We construct different datasets to evaluate our approach. To evaluate the performance of an approach in detecting Trapdoor tokens, we use the ground-truth dataset collected in Section~\ref{sec:labelling}, which consists of 18,548 non-Trapdoor tokens and 11,943 Trapdoor tokens. In addition, we also want to verify if the proposed approach is able to detect five techniques mentioned in Section~\ref{subsec:Trapdoor_definition} in a Trapdoor contract. To do that, we first retrieve all Trapdoor instances in the ground truth dataset and construct five experimental datasets based on the \duy{Trapdoor indicator} list in each token contract. For example, if a Trapdoor token A contains three different techniques: AL, EP, and ES, it will be labelled as ``positive'' (label = 1) in AL, EP, and ES datasets and as ``negative'' (label = 0) in FM and IC datasets. As a result, the class distribution of five separated datasets 
is depicted in Table~\ref{tab:type_distributions}.

\vspace{-5pt}
\begin{table}[htb!]
\centering
\caption{Five Trapdoor types dataset.}
\label{tab:type_distributions}
\begin{tabular}{|c|c|c|}
\hline
\textbf{Types}      &   \textbf{Negative }   &   \textbf{Positive}          \\ \hline
 Exchange Permission (EP) & 394 & 11,549 \\
 Exchange Suspension (ES) & 4,155 & 7,788 \\
 Amount Limit (AL) & 4,044 & 7,899 \\
 Fee Manipulation (FM) &6,719 & 5,224\\
 Invalid Callback (IC) & 11,712 & 231\\
\hline
\end{tabular}
\end{table}

\textbf{Classification models:} Our goal is to train and test a few well-known machine classification methods using our proposed dataset and compare their performance to find the most suitable classification model for this problem. These models are KNN~\cite{cover1967nearest}, SVM~\cite{hearst1998support}, Random Forest (RF)~\cite{svetnik2003random}, XGBoost (XGB)~\cite{chen2016xgboost}, and LightGBM (LGBM)~\cite{ke2017lightgbm}. 

\textbf{Experimental settings:}
In our experiment, instances in the dataset are randomly shuffled and split into a training set (80\%) and a test set (20\%). While the training set is used to train our detection models, a test set is used to validate the model's performance. During the model training process, we adopt the 10-fold cross-validation to train and validate the selected model on the training set across different sets of parameters to figure out the optimal model. The best model was used to classify the tokens in the unseen test dataset. The parameter sets for each model are depicted in Table~\ref{tab:model_parameters}. To make our results more reliable, we repeated the experiment process ten times, and the final result was obtained by taking the average. 

\begin{table}[htb!]
\centering
\caption{Detection model parameter settings. The highlighted values show the best settings achieved from our experiments.}
\label{tab:model_parameters}
\begin{tabular}{|c|l|c|}
\hline
\multicolumn{1}{|c|}{\textbf{Classifier}} & \multicolumn{1}{c|}{\textbf{Parameters}} & \multicolumn{1}{c|}{\textbf{Total}} \\ \hline
\multirow{2}{*}{KNN}                      & n\_neighbors = {[}\textbf{5}, 10, 15{]}           & \multirow{2}{*}{9}                                 \\
                                          & leaf\_size: {[}\textbf{10}, 50, 100{]}            &                                                    \\\hline
\multirow{2}{*}{SVM}                      & kernel = {[}linear, \textbf{poly}{]}               & \multirow{2}{*}{5}                                 \\
                                          & degree = {[}2, \textbf{3}, 4, 5{]} (kernel=poly)                       &                                                    \\\hline
\multirow{2}{*}{RF}                       & n\_estimators = {[}\textbf{50}, 100, 200{]}       & \multirow{2}{*}{9}                                 \\
                                          & min samples leaf = {[}\textbf{5}, 10, 50{]}       &                                                    \\\hline
\multirow{2}{*}{XGB}                      & learning\_rate = {[}\textbf{0.1}, 0.2, 0.5{]}     & \multirow{2}{*}{9}                                 \\
                                          & n\_estimators = {[}50, 100, \textbf{500}{]}       &                                                    \\\hline
\multirow{2}{*}{LGBM}                     & learning\_rate = {[}0.1, \textbf{0.2}, 0.5{]}     & \multirow{2}{*}{9}                                 \\
                                          & n\_estimators = {[}50, 100, \textbf{500}{]}       &  \\                                                 \hline
\end{tabular}
\end{table}


\textbf{Evaluation Metrics:} Every selected classifier's performance is evaluated by using four different standard metrics: \texttt{Accuracy}, \texttt{Precision}, \texttt{Recall}, and \texttt{F1-score}. These metrics are calculated based on the numbers of true positives (\texttt{TP}), true negatives (\texttt{TN}), false positives (\texttt{FP}), and false negatives (\texttt{FN}) in the detection result. $\textbf{Accuracy}\triangleq (\texttt{TP}\hspace{-1pt} +\hspace{-1pt} \texttt{TN})/(\texttt{TP} \hspace{-2pt}+\hspace{-2pt} \texttt{FP} \hspace{-2pt}+\hspace{-2pt} \texttt{TN} \hspace{-2pt}+\hspace{-2pt} \texttt{FN})$ is the fraction of correct predictions, $\textbf{Precision}\triangleq \texttt{TP}/(\texttt{TP + FP})$ is the fraction of the actual Ponzi applications out of all the predicted Ponzi by the method, $\textbf{Recall} \triangleq \texttt{TP}/(\texttt{TP + FN})$ is the fraction of detected scams among all actual scams, and $\textbf{F1-score} \triangleq (2\cdot\texttt{Precision}\cdot\texttt{Recall})/(\texttt{Precision + Recall})$ is the harmonic mean of \texttt{Precision} and \texttt{Recall}.

\subsection{Experimental results}
\textbf{Feature types and detection models evaluation.}
A comparison of performances between detection models listed above with two different kinds of features is summarised in Table~\ref{tab:experiment_results}. The best scores are shown in bold font. Overall, models using code features perform better in all metrics than those using exchange features. We expected that the exchange features would perform better than code features because Trapdoor tokens have different behaviours, while a contract source code may be obfuscated by creators. In contrast, the code feature is very efficient in detecting Trapdoor tokens with all metrics over 80\% for all machine learning models except SVM. Although exchange-based models have lower performance than opcode-based models, their performances are still high in some tree-based models such as RF, XGB, and LGBM with F1-scores over 92\%. The possible reason of why the exchange features are less efficient than code features may be that some tokens are underperforming tokens. These tokens have a very short lifetime or a poor exchange history with buy transactions only. Another reason may be that some Trapdoor tokens have an active trading history because they let users exchange back and forth freely at the beginning to attract more users before suspending trading or updating extreme exchange fees. Those reasons will make tokens misclassified from their behaviours.


 As shown in Table~\ref{tab:experiment_results}, the SVM model had the lowest performance among proposed models when combined with exchange features. However, this model yielded higher performance when it used code features. From this table, we also observed that tree-based classifiers were more efficient in Ponzi detection than other algorithms, regardless of feature kind. Three models, RF, XGB, and LGBM, achieve very high performance, specifically over 94\% for all metrics with code features and over 90\% for all metrics with exchange features.  Among tree-based models, LGBM achieved the best performance. As we mentioned in the experiment setting, our models adopted different parameters, and their performance is measured to figure out the optimal combination of those parameters. As a result, the best combinations are 
 highlighted in Table~\ref{tab:model_parameters}.



\begin{table*}[ht]
\centering
\caption{Comparison of different detection models for Trapdoor tokens.}
\label{tab:experiment_results}
\begin{tabular}{|l|cc|cc|cc|cc|}
\hline
\multirow{2}{*}{\textbf{Model}} & \multicolumn{2}{l|}{\textbf{Accuracy}} & \multicolumn{2}{l|}{\textbf{Precision}} & \multicolumn{2}{l|}{\textbf{Recall}} & \multicolumn{2}{l|}{\textbf{F1-Score}} \\ \cline{2-9} 
                                     & \textbf{Exchange}   & \textbf{Code}    & \textbf{Exchange}    & \textbf{Code}    & \textbf{Exchange}  & \textbf{Code}   & \textbf{Exchange}   & \textbf{Code}    \\ \hline
KNN                                  & 0.790               & 0.876            & 0.756                & 0.860            & 0.680               & 0.808           & 0.716               & 0.833            \\
SVM                                  & 0.705               & 0.840           & 0.934                & 0.959            & 0.260              & 0.606           & 0.407               & 0.687            \\
RF                                   & 0.941               & 0.967            & 0.944               & 0.970            & 0.903              & 0.944           & 0.923               & 0.957           \\
XGB                                  & 0.952               & 0.974            & 0.945                & 0.971            & 0.931              & 0.961          & 0.938               & 0.966            \\
\textbf{LGBM}                                 & \textbf{0.957}      & \textbf{0.976}   & \textbf{0.953}       & \textbf{0.975}   & \textbf{0.935}     & \textbf{0.963}  & \textbf{0.944}      & \textbf{0.969}   \\ \hline
\end{tabular}
\end{table*}

\begin{figure}[ht]
\centering
\includegraphics[width=0.4\textwidth]{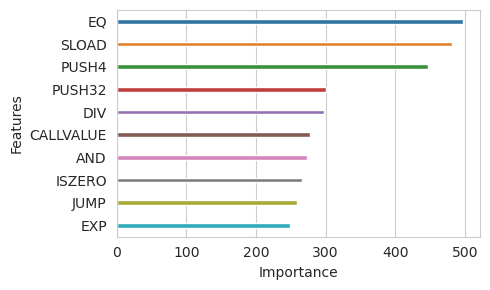}
\caption{The importance of the 10 most significant features.}
\label{fig:top_ten_importance_features}
\end{figure}

\textbf{Important features analysis.} 
 To understand the effectiveness of opcode features in detecting Trapdoor tokens and make an effort to interpret the results, 
 we retrieved the \textit{importance} of each opcode from the best performance model LGBM in the previous experiment. The importance of a feature is defined by the number of times this feature is used to split the data across all decision trees. Hence, the most important feature is the most efficient opcode used to discriminate non-Trapdoor and Trapdoor tokens. To this end, we extracted the top ten important features (Fig.~\ref{fig:top_ten_importance_features}) and counted their occurrences in each token in our dataset. The statistics and description of those opcodes are provided in Table~\ref{tab:feature_analysis}.

According to Table~\ref{tab:feature_analysis}, we can clearly observe that the top important opcodes appear in the Trapdoor contracts more than non-Trapdoor contracts. This is not surprising since their combinations can indicate popular Trapdoor techniques. For instance, the combinations of data load operations (e.g., \texttt{SLOAD}, \texttt{CALLVALUE}) and some logical operations such as \texttt{EQ}, \texttt{ISZERO}, and \texttt{AND} show the technique of using comparisons as the condition to throw an exception intentionally, such as comparing a transfer amount and a transfer limit (\texttt{SLOAD}, \texttt{CALLVALUE}) or checking whether a seller's address is in a whitelist for exchange permission (\texttt{SLOAD}, \texttt{ISZERO}). Another example is the combinations of \texttt{CALLVALUE}, \texttt{PUSH} and some calculation opcodes such as \texttt{DIV} and \texttt{EXP}. These combinations may show that loading a transfer amount and applying a transfer fee occur more frequently in Trapdoor tokens.

\begin{table}[htb!]
\centering
\caption{Average frequencies of top ten important opcode features in Trapdoor and non-Trapdoor tokens.}
\label{tab:feature_analysis}
\begin{tabular}{|c|c|c|}
\hline
\textbf{Opcode}&  \textbf{Trapdoor}  & \textbf{non-Trapdoor} \\ \hline
EQ  &51 & 20 \\  
SLOAD     &  78 & 25 \\   
PUSH4     & 52 & 23 \\  
PUSH32    & 34 & 14 \\  
DIV  & 21 & 10 \\  
CALLVALUE  &  22 & 5 \\  
AND & 152 & 72 \\  
ISZERO & 111 & 43 \\  
JUMP  & 245 & 85 \\  
EXP &18 & 7 \\  
\hline
\end{tabular}
\end{table}


\textbf{Detecting Trapdoor injected techniques.} 
As shown in Table~\ref{tab:type_distributions}, the difference between the numbers of positive samples and negative samples is huge and far from the 1:1 ratio in all five given datasets. The class imbalance problem may impact the detection performance of the machine learning models because an imbalanced-dataset-trained model makes it hard to discover patterns for the minority class and overwhelms the majority class. To solve this problem, we applied data sampling techniques to balance our datasets. We adopted the oversampling method SMOTE~\cite{han2005borderline} to generate new instances for minority class by cloning minority samples which have more than half of the K nearest neighbours being majority samples. In data sampling, we used the default setting for SMOTE with k = 5. With the SMOTE algorithm, the training datasets will have the same number of positive and negative samples that are used for the forthcoming model's training. The detection results for each type of Trapdoors are shown in Table~\ref{tab:trapdoor_technique_detection}.  Generally, all models perform very well in detecting a Trapdoor technique embedded into a token contract with code features. Similarly to the previous experiment, tree-based detection models yielded a higher value in all metrics than KNN and SVM. In this experiment, XGB and LGBM achieved better performances than RF, and LGBM is the best model for this detection problem.

\begin{table}[ht]
\centering
\caption{Trapdoor technique detection results.}
\label{tab:trapdoor_technique_detection}
\resizebox{0.5\textwidth}{!}{
\begin{tabular}{|c|c|c|c|c|c|}
\hline
\textbf{Type}       & \textbf{Model} & \textbf{Accuracy} & \textbf{Precision} & \textbf{Recall} & \textbf{F1-Score} \\ \hline
\multirow{5}{*}{EP} & KNN            & 0.890             & 0.918              & 0.856           & 0.885            \\
                    & SVM            & 0.884             & 0.965              & 0.776           & 0.822            \\
                    & RF             & 0.960             & 0.974              & 0.956           & 0.964             \\
                    & XGB            & 0.968             & 0.974              &  0.968          & 0.971    \\
                    & \textbf{LGBM}           & \textbf{0.970}             & \textbf{0.977 }             & \textbf{0.970 }          & \textbf{0.973 }           \\ \hline
\multirow{5}{*}{ES} & KNN            & 0.891             & 0.924              & 0.862           & 0.0.892             \\
                    & SVM            & 0.895             & 0.962              & 0.811           & 0.848             \\
                    & RF             & 0.960             & 0.970              & 0.961           & 0.966             \\
                    & XGB            & 0.966             & 0.972              & 0.970           & 0.971    \\
                    & \textbf{LGBM}           & \textbf{0.968}             & \textbf{0.974}              & \textbf{0.972}           & \textbf{0.973}             \\ \hline
\multirow{5}{*}{AL} & KNN            & 0.889             & 0.895              & 0.841           & 0.867             \\
                    & SVM            & 0.879             & 0.959              & 0.730           & 0.781             \\
                    & RF             & 0.969             & 0.970              & 0.959           & 0.964             \\
                    & XGB            & 0.972             & 0.972              & 0.968           & 0.970    \\
                    & \textbf{LGBM}           & \textbf{0.974}             & \textbf{0.975}              & \textbf{0.970}           & \textbf{0.972}             \\ \hline
\multirow{5}{*}{FM} & KNN            & 0.906             & 0.931              & 0.884           & 0.906             \\
                    & SVM            & 0.909             & 0.963              & 0.840           & 0.870             \\
                    & RF             & 0.964             & 0.971              & 0.966           & 0.968             \\
                    & XGB            & 0.970             & 0.973              & 0.974           & 0.973    \\
                    & \textbf{LGBM}           & \textbf{0.971}             & \textbf{0.975}              & \textbf{0.974}           & \textbf{0.975}             \\ \hline
\multirow{5}{*}{IC} & KNN            & 0.917             & 0.882              & 0.876           & 0.874             \\
                    & SVM            & 0.922             & 0.960              & 0.851           & 0.878             \\
                    & RF             & 0.969             & 0.972              & 0.965           & 0.968             \\
                    & XGB            & 0.974             & 0.971              & 0.971           & 0.971    \\
                    & \textbf{LGBM}           & \textbf{0.975}             & \textbf{0.972 }             & \textbf{0.972}         & \textbf{0.972 }            \\ \hline
\end{tabular}}
\end{table}


\section{Trapdoor Analysis}
\label{sec:trapdoor_analysis}
In this section, we analyse the collected Trapdoor tokens from different perspectives to obtain more insights into this scam.
\subsection{General Overview}

\begin{figure}[ht]
\centering
\includegraphics[width=0.4\textwidth]{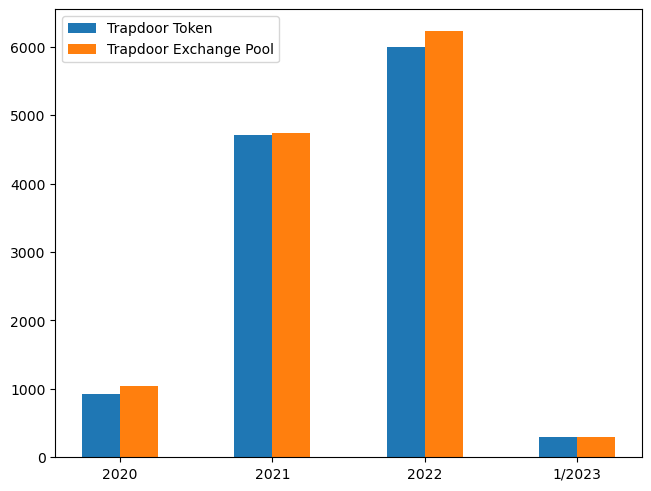}
\caption{The scale of Trapdoor tokens on Uniswap.}
\label{fig:trapdoor_token}
\vspace{-10pt}
\end{figure}

\begin{figure}[ht]
\centering
\includegraphics[width=0.4\textwidth]{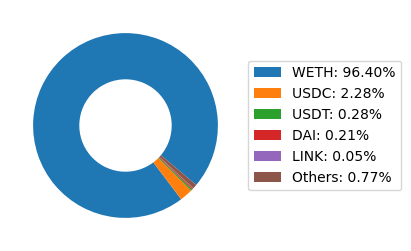}
\caption{Top 5 tokens paired with Trapdoor tokens.}
\label{fig:top_paird_token}
\vspace{-5pt}
\end{figure}

\textbf{The scale of Trapdoor tokens.} As depicted in the Fig.~\ref{fig:trapdoor_token}, the creation of Trapdoor tokens follows the overall trend of UniswapV2 (see Fig.~\ref{fig:token_and_exchange_pool}). Surprisingly, the Trapdoor scam is not new. It started appearing on UniswapV2 after roughly two weeks since this platform launched. Trapdoor tokens grew 
significantly in the second half of 2020, accounting for 4\% of all tokens listed in this year. The proportion of Trapdoor tokens increased by 11\% in 2021 and dropped to 8.5\% in 2022. The same phenomenon also occurs in the creation of trapdoor exchange pools. Generally, 99\% Trapdoor tokens have only one corresponding exchange pool on the UniswapV2, and these tokens often pair with high-value tokens (Section~\ref{sec:Trapdoor_token}). Particularly, we collected 87 different high-value tokens that were paired with Trapdoor tokens across 12,317 different exchange pools. The top 5 tokens paired the most with Trapdoor tokens are WETH (96,40\%), USDC (2,28\%), USDT (0,28\%), DAI (0,21\%), and UNI (0,05\%) (see Fig.~\ref{fig:top_paird_token}).

\textbf{Trapdoor lifetime.} One of Blockchain's characteristics is immutability, which means once a token is deployed onto Ether-eum, it will be there forever. Thus, we define the \textit{lifetime} of a token as follows: The lifetime of a token starts at the block where this token was created and finishes at the block where the last \texttt{Transfer} event of this token was emitted. We found that 7,282 tokens (60.9\%) in our dataset have a lifetime of less than 24 hours. In particular, 2,918 tokens (24.4\%) had a lifetime of even less than 1 hour, but they were still exchanged by users. For example, the token \texttt{EternalMoon}~\cite{EternalMoon} was purchased by 132 different users, although its \texttt{lifetime} only lasted for 68 blocks. The remaining 39.0\% are Trapdoor tokens that seem to be more hidden and difficult to detect. Surprisingly, 144 tokens (1.2\%) among them lived longer than one year. The most successful long-life scam token is \texttt{Mommy Milkers}~\cite{MommyMilkers}, which stole 180 WETHs (US\$ 397,512) from about 1,400 investors.

\subsection{Scammer Tactics}

\begin{figure}[ht]
\centering
\includegraphics[width=0.3\textwidth]{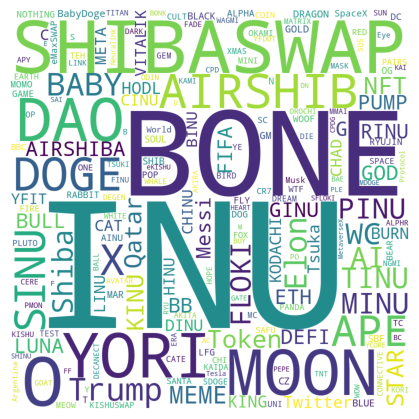}
\caption{A symbol cloud of Trapdoor tokens.}
\label{fig:symbol_wordcloud}
\end{figure}

\textbf{Fake tokens.}
While tokens should have a unique name and symbol, scammers can still name their Trapdoor token similarly to a popular token to make users mistakenly think that they are exchanging with the original one. Fig.~\ref{fig:symbol_wordcloud} shows the word cloud graph of symbols in our Trapdoor tokens set. The font size of each word represents the number of tokens that have named its symbol with this word. By comparing token names and symbols, we found 959 tokens (8.0\%) that have the same name or symbol as high-value tokens.  For instance, 66 different Trapdoor tokens have the same symbol or name as the high-value token \texttt{Bone ShibaSwap (BONE)}\footnote{https://coinmarketcap.com/currencies/bone-shibaswap/} and those tokens have successfully scammed 8,786 WETH from 5,754 investors in total. This strategy was also observed in an earlier work by Xia \et~\cite{Xia:2021} for the general Rug-pull scams.

\textbf{Scam token cloning.} 
We grouped Trapdoor tokens if their contracts are the same (100\% similarity) and found that 874 groups were constructed from 11,943 Trapdoor tokens. Table~\ref{tab:tokensimilar} presents more information about the five largest token groups. 
The largest group contains 2,417 different tokens, while the smallest group contains only one instance. The largest group contains different tokens which have the same source code as the token \texttt{CSTs}. Those tokens were created by 1,317 different accounts. Furthermore, the second largest token group contains 1,971 same source-code tokens created by 1,894 different accounts. Looking carefully at the table, we can observe that a token is created by a distinct account in most of the groups. For instance, except for the biggest group, about 95\% of tokens in the group have been created by different accounts. 
From the scammer's perspective, we speculate that using multiple accounts to deploy different clones of the same scam tokens will make the scams look unrelated and harder to trace, e.g. by detection tools relying on the guilt-by-association rule (e.g. in Xia~\et~\cite{Xia:2021}). Generating a large number of scam tokens from the same or a few original tokens seems also a cheap way to catch more investments (e.g. from auto-trading bots~\cite{cernera2023token,Cernera_etal_WWWCompanion2023}).  

\begin{table}[htb!]
\centering
\caption{Top-five token groups, each has the same source code across all tokens.}
\label{tab:tokensimilar}

\begin{tabular}{|c|c|c|}
\hline
\textbf{Representative token}      &   \textbf{Tokens}   &   \textbf{Creators}          \\ \hline
CSTs (CSTs)~\cite{CSTs} & 2,417 & 1,317 \\
Gaifu Kaisei (Kakkazan)~\cite{Kakkazan} & 1,971 & 1,894 \\
Fifth Dimension (5D)~\cite{5D} & 1,076 & 1,051 \\
Okamikosis (OKAMI)~\cite{OKAMI} & 931 & 925 \\
NFTvft (NFTvft)~\cite{NFTvft} & 859 & 809 \\
\hline
\end{tabular}
\end{table}
\vspace{-10pt}

\subsection{Trapdoor Impacts}
To evaluate 
the financial impact of this scam, we first collect all swap events from exchange pools of a token. Then, the profit of each Trapdoor token is calculated by subtraction the total amount of high-value tokens sent to the exchange pool and the total amount of that token withdrawn by users from a pool. Moreover, we also convert all profit into US dollars (USD) according to the price of the corresponding high-value token at the time a Trapdoor token was created. 
The analysis was conducted on 11,943 Trapdoor tokens, and their economic statistics are shown in Table~\ref{tab:trapdoor_impacts}, with the top five profit tokens identified. In total, the scam tokens in our dataset have collected over 1.1 billion US dollars from 267,809 unique investor addresses. Among them, the MYOBU token has gained a total of US\$ 8.2 million from 3,620 investors.

\begin{table}[htb!]
\centering
\caption{Five most profitable tokens \& number of investors.}
\label{tab:trapdoor_impacts}
\begin{tabular}{|c|c|c|}
\hline
\textbf{Token}&\textbf{Profit (USD)}&\textbf{Investors} \\\hline
 Myōbu (MYOBU)~\cite{MYOBU}&   \$8,237,866 & 3,620\\
 FinXFlo.com (FXF)~\cite{FXF}&  \$6,847,401 & 381\\
 WallStreetBets (WSB)~\cite{WSB}&  \$6,656,059 & 323\\
 corion.io (CORX)~\cite{CORX}& \$5,281,929 & 410\\
 Hashmasks (HM)~\cite{HM}& \$4,916,379 & 207\\
 \hline \textbf{Total} (11,943 tokens) &\$1,159,857,503 & 267,809\\
\hline
\end{tabular}
\end{table}
\vspace{-10pt}

\section{Related work}
\label{sec:related_work}
Trapdoor is a new kind of fraud combining intentional 
programming bugs 
with malicious behaviours. In recent years, this kind of scam has drawn 
a lot of attention 
from the blockchain community. 
There have been a number of blog posts~\cite{coinbrain_post,cryptobriefing_post,quillaudits_post,rampiro_post} about scams related to Trapdoor, along with the tips on how to identify them, written by technology enthusiasts or by the DEXs themselves. While such posts provide some good high-level descriptions of the scams to raise awareness, most of them provide no comprehensive analysis of the tactics employed by the scammer. An example is the blog post of quillaudits~\cite{quillaudits_post}. Although it provides an instruction for analysing scam tokens with some code analysis, the instruction does not cover all techniques used by the scammers. Moreover, the analysis and explanation for the examples in the post are quite minimal and inadequate for novice investors.

Several methods have been proposed for detecting scams and security issues on blockchain based on smart contracts. Bartoletti \emph{et al.}~\cite{BARTOLETTI2020259} proposed four criteria to detect Ponzi schemes by inspecting their Solidity source codes manually. Chen \emph{et al.}~\cite{chen2018detecting,chen2019exploiting} proposed machine-learning-based detection tools for detecting Ponzi smart contracts based on opcode features where their Solidity source code is unavailable. Many other kinds of blockchain scams have been researched apart from Ponzi schemes such as 
ICO scams~\cite{liebau2019crypto,zetzsche2017ico}, phishing scams~\cite{chen2020phishing,wu2020phishers}, 
and rug pulls~\cite{cernera2023token,mazorra2022not,nguyen2023rug,Xia:2021}. Despite this, the research on Trapdoor tokens has not yet been systematically studied, and existing techniques cannot be applied to identify Trapdoor techniques directly. There are some relevant papers~\cite{mazorra2022not,nguyen2023rug,Xia:2021} which briefly mention Trapdoor as a special trick of Rug Pull. However, their discussion revolves around a simple demonstration of a Trapdoor without any further analysis of trapping techniques, taxonomy, or economic impact. 

Additionally, there are some online tools developed by individuals or companies such as ``honeypot.is''~\cite{honeypotis}, ``detecthoneypot.com''~\cite{detecthoneypot}, Token Sniffer~\cite{tokensniffer}, and GoPlus~\cite{goplus}. Although the first two tools work primarily on the detection of Trapdoor/Honeypot tokens, their drawbacks are apparent. First, if the tool cannot simulate a buy, it cannot detect if a given token is a scam. Failure to buy could be due to many reasons, including empty liquidity in a trading pool or a scammer turning off the trading mode (preventing both buying and selling) to trap all participants' funds. Moreover, these tools do not provide any further explanation for their detection outcome. Thus, the users gain little in their knowledge of scam techniques and maneuvers using such tools. The tool provided by ``tokensniffer.com'' lists more token audit details than the others. This tool has been quite a success and was even cited by the US Department of the Treasure~\cite{us_department} in testimony before the Senate Banking Committee~\cite{us_banking_committee}. GoPlus is a commercial tool that performs token auditing based on token contracts to check security issues. This tool also provides audit APIs used by popular DEXs and crypto analytic platforms. 
These tools analyse a given contract from various perspectives and provide an audit score regarding a list of criteria that helps investors estimate the risk before making investment decisions. However, these tools assume that investors or users already have sufficient knowledge about blockchain and DEXs. Last but not least, the criteria listed by these tools are quite general and cannot cover all cases of Trapdoor. Exposing all the criteria also leads to the risk that they will be bypassed by more sophisticated scam tokens one day. To the best of our knowledge, this paper is the first work that considers, analyses, discusses and detects Trapdoor scams. Our work also proposes a taxonomy, a study dataset, and an automated tool for analysing and studying this kind of scam.

\section{Conclusion}
\label{sec:conclusion}

In this work, we provided a comprehensive analysis of Trapdoor tokens, a new emerging financial fraud on Uniswap, one of the most popular decentralised exchanges. Our study revealed a range of simple to advanced techniques successfully employed by scammers to trap 
investment funds. \hg{Moreover, we developed 
an effective tool called TrapdoorAnalyser that can accurately detect and label Trapdoor tokens listed on Uniswap as long as their source codes are available. 
Extensive evaluations showed that TrapdoorAnalyser outperform GoPlus, a comparable commercial tool from the industry. 
Thanks to the large Trapdoor/non-Trapdoor dataset constructed based on TrapdoorAnalyser, we trained machine learning models that can detect even Trapdoor tokens with no available source codes with very high accuracy. 
The verified dataset of Trapdoor scam tokens, the machine learning detection models, together with the detailed analysis provided in this paper will facilitate future studies into crypto scams, which are currently running rampant across different investment platforms. In future work, we plan to extend our approaches to other DEXs and blockchain platforms to discover new Trapdoor variants.} 


\section*{Acknowledgements}
This work was supported by the Australia Research Council Discovery Project Grant DP200100731.
\bibliographystyle{elsarticle-num-names} 
\bibliography{main}






\end{document}